\documentstyle[11pt,epsf]{article}
\newcommand{\al}{\alpha}
\newcommand{\ap}{\simeq}
\newcommand{\bt}{\beta}
\newcommand{\de}{\delta}
\newcommand{\ch}{\chi}
\newcommand{\De}{\Delta}
\newcommand{\ep}{\epsilon}
\newcommand{\et}{\eta}
\newcommand{\ev}{\end{verbatim}} 
\newcommand{\fr}{\frac}
\newcommand{\Ga}{\Gamma}
\newcommand{\gm}{\gamma}
\newcommand{\hb}{\hbar}

\newcommand{\lb}{\label}

\newcommand{\lf}{\left}
\newcommand{\lm}{\lambda}

\newcommand{\na}{\nabla}
\newcommand{\Om}{\Omega}
\newcommand{\om}{\omega}
\newcommand{\ph}{\phi}
\newcommand{\ps}{\psi}
\newcommand{\rh}{\rho}
\newcommand{\rt}{\right}
\newcommand{\Si}{\Sigma}
\newcommand{\si}{\sigma}
\newcommand{\ta}{\tau}

\newcommand{\te}{\theta}

\newcommand{\vb}{\verb}

\newcommand{\be}{\begin{equation}}
\newcommand{\ee}{\end{equation}} 
\newcommand{\eei}{\end{equation}\indent\indent}
\newcommand{\bc}{\begin{center}}
\newcommand{\ec}{\end{center}}
\newcommand{\ber}{\begin{eqnarray}}
\newcommand{\ear}{\end{eqnarray}}
\newcommand{\ba}{\begin{array}}
\newcommand{\ea}{\end{array}}

\newcommand{\nn}{\nonumber\\}

\newcommand{\p}{\partial}

\def\case#1/#2{\textstyle\frac{#1}{#2} }
\begin{document}
\title{Quantum Perturbative Approach to Discrete Redshift.}
\author{
         Mark D. Roberts,\\
         Department of Mathematics and Applied Mathematics,\\  
University of Cape Town,\\
Rondebosch 7701,\\
South Africa.\\
roberts@cosmology.mth.uct.ac.za\\
www.cosmology.mth.uct.ac.za/$\sim$~roberts\\\\
and\\
California Institute for Physics and Astrophysics,\\
366 Cambridge Ave.,  Palo Alto, CA 94306.} 
\date{\today}
\maketitle
\vspace{0.1truein}
\bc Report Number:  uct-cosmology/00/01 \ec
\bc Eprint number:  astro-ph/0002434\ec
\bc Comments:  53 pages  167372  bytes, 120 references,
no diagrams,  LaTex2e.\ec
\bc Keywords:  Cosmology:~~  theory-galaxies:~~  distance and redshifts.\ec
\bc 1999 PACS Classification Scheme:\ec
\bc http://publish.aps.org/eprint/gateway/pacslist \ec
\bc 98.80-k,03.65-w\ec
\bc 1991 Mathematics Subject Classification:\ec
\bc http://www.ams.org/msc \ec
\bc 83F05, 81S99,  85A40.\ec
\newpage
\begin{abstract}
On the largest scales there is evidence of discrete structure,
examples of this are superclusters and voids and also by redshift 
taking discrete values.   In this paper it is proposed that discrete 
redshift can be explained by using the spherical harmonic 
integer $l$;  this occurs both in the metric or density perturbations and 
also in the solution of wave equations in Robertson-Walker spacetime.   
It is argued that the near conservation of energy implies that 
$l$ varies regularly for wave equations in Robertson-Walker spacetime,  
whereas for density perturbations $l$ cannot vary regularly.   
Once this is assumed then perhaps the observed value of discrete redshift 
provides the only observational or experimental data that directly requires an 
explanation using both gravitational and quantum theory.   
In principle a model using this data could predict the scale factor $R$  
(or equivalently the deceleration parameter $q$).    
Solutions of the Klein-Gordon equation in Robertson-Walker 
spacetimes are used to devise models which have redshift taking discrete
values,  but they predict a microscopic value for $R$.   A model in which
the stress of the Klein-Gordon equation induces a metrical perturbation of 
Robertson-Walker spacetime is devised.  Calculations based upon this model 
predict that the Universe is closed with $2q_0-1=10^{-4}$.
\end{abstract}
{\small\tableofcontents}
\section{Introduction}
\label{sec:intro}
\subsection{evidence of discrete structure}
There are at least {\sc four} types of evidence of discrete structure on the
largest scales.   
The {\sc first} example of discrete structure is given by discrete redshift,  
the evidence for this will be discussed in the next paragraph.  
Redshift comes in discrete values with the characteristic velocity
$v_{I}=72.2\pm0.2~{\rm Km.}~{\rm s}^{-1}$,  and this leads to a characteristic 
length $l_{dr}=v_{I}H_{0}^{-1}=3\pm0.8.10^{22}$ meters,  and a characteristic 
period $t_{dr}= v_{I}H_{0}^{-1}c^{-1}=3.2\pm0.8.10^{6}$ years.   
Here this result is used in preference to the other examples of discrete 
structure because of the advantage of having a qualitative result,  
namely $v_I=72.2\pm0.2~{\rm Km.s.}^{-1}$;  the actual techniques used
might be applicable to the other cases.
A {\sc second} example is that the Universe appears to 
consist of superclusters and voids  
Saunders {\it et al} (1991) \cite{bi:sau} recent studies of clustering can
be found in Cohen (1999) \cite{bi:cohen},  superclusters and voids occur with 
apparent regularity  (Broadhurst {\it et al} (1990) 
\cite{bi:BEKS}).   Assuming $q_{0}=\fr{1}{2}$ they have a characteristic 
scale of $l_{sv}= 128~{\rm h}^{-1}{\rm Mpc}=5.3\pm2.6.10^{24}$  meters,  and 
hence a characteristic period of $t_{sv}=\fr{l_{sv}}{c}=5.6\pm10^{8}$ years.
Tytler {\it et al} (1993) \cite{bi:TSK} do not confirm Broadhurst {\it et al's}
result:  instead of Broadhurst {\it et al's} "apparent regularity with a scale 
of $ 128 {\rm h}^{-1}{\rm Mpc.}$" they find "There is no significant 
periodicity 
on any scale from $10$ to $210 {\rm h}^{-1}{\rm Mpc.}$"   The result was 
also looked at by Willmer {\it et al} (1994) \cite{bi:WKS}.
Einasto {\it et al} (1997) \cite{bi:EE} present evidence for a quiasiregular
three-dimensional network of rich superclusters and voids,
with regions separated by $\sim$120Mpc.;  
and they say that ``if this reflects the distribution of all matter,
then there must exist some hitherto unknown process 
that produces regular structure on large scales''.
A {\sc third} example of discrete structure is that some normal elliptical 
galaxies have giant shells surrounding them,  
Malin and Carter (1980) \cite{bi:MC}.  
These shells probably consist of stars,  the 
most likely method of their formation is from an intergalactic shock wave or an
explosive event in the galaxy.  
The {\sc fourth} example is of discrete properties from geological
time scales.
The characteristic periods $t_{dr}$ and $t_{sv}$  
are larger than typical geological frequencies.   
For example 
Kortenkamp and Dermott (1998) \cite{bi:KS} give a periodicity of $t_d=10^5$ 
years for the accretion rate of interplanetary dust.   The rate of accretion
of dust by the Earth has varied by a factor of 2 or 3.   
Extraterrestrial helium-3 concentrations in deep sea cores display 
a similar periodicity but are $5.10^4$ years out of phase.
The magnetic polarity time sequence 
(p.672 of Larson and Birkland (1982) \cite{bi:LB}) 
gives reversals in the earths magnetic field occurring at intervals 
$t_{m}=10^{3}~~{\rm to}~~5.10^{3}$ years;  
and the period of recent glacier advance and retreat is about 
$t_{g}=2.10^{3}$  years,
(p.488 Larson and Birkland (1982) \cite{bi:LB}).  
Naidu and Malmgren (1995) \cite{bi:NM} give a periodicity of $t_m=2,200$ years
for the Asian monsoon system,  this is found by looking at fluctuations in
upwelling intensity in the western Arabian sea.
Clearly it would be good if there was an 
explanation for these characteristic periods and even better if they could 
be used to predict presently unknown quantities.  The obvious factor to try
to predict is the size of the Universe as given by the scale factor $R$,  
assuming a Robertson-Walker Universe this is equivalent to deriving a value of
the deceleration parameter $q$.   
Now the value of the deceleration parameter is
assumed in the derivation of $t_{sv}$,  thus models using the characteristic 
period of discrete redshift are studied.
\subsection{observations of discrete red shift}
\label{sec:obs}
Perhaps the first paper advocating that redshift can only occur in specific
discrete values was Cowan (1969) \cite{bi:cowan} who found a periodic 
clustering of redshifts.   This was confirmed by Karlsson (1971) 
\cite{bi:karlsson} who found a number of new peaks in the distribution of
quasi-stellar objects;  these,  together with the peaks at $z=1.956$ and
$z=0.061$ formed a geometric series:  $z_1=1.96,  z_2=1.41,  z_3=0.96,  
z_4=0.060,  z_5=0.30,  z_6=0.06$;
this was supported by Arp {\it et al} (1990) \cite{bi:ABCZ}.
Green and Richstone (1976) \cite{bi:GR} did a search for peaks and 
periodicities in the redshift distribution of a sample of quasars and 
emission-line galaxies independent of that used in earlier work.
In agreement with the results of Burbidge and O'Dell (1972) \cite{bi:BOD},  no 
statistically significant peak was found at a redshift of $1.95$,  nor any 
significant periodicity in redshift in either the sample of quasars alone or 
the sample of quasars and galaxies together.   The strong spectral power peak 
in their distribution of galaxy redshifts,  estimated at a confidence level 
of $97.5$ percent,  is completely absent in Green and Richstone's analysis;  
they conclude that the observed redshift distribution is consistent with a
random sample of discrete values from a smooth,  
aperiodic underlying population.
Tifft (1976) \cite{bi:tifft76} claimed  
that well known local galaxies,  especially {\sf M31},  
were claimed to consist of two basically opposed streams 
of outflow material which have an intrinsic difference in red shift of 
$70-75~ {\rm Km}.~ {\rm s}^{-1}$.  
Tiffts' result was questioned by Monnet and Deharneny (1977) \cite{bi:MD} 
who say that the two opposing streams of material suggest that the best galaxy 
candidate for a direct test of Tiffts' result would be a near face-on galaxy:
the Doppler effect due to the rotation is minimized and any expansive motion
in the spiral arms,  as connected with the uneven distribution of neutral gas,
which can be of the order of $10-25~ {\rm Km.s.}^{-1}$ at most,  gives 
entirely negligible Doppler effect.   On the other hand a real intrinsic 
redshift would still exhibit its full $70-75~ {\rm Km.s.}^{-1}$ discontinuity.
Monnet and Deharneny chose the nearly face-on galaxy NGC628 and find no
intrinsic effects as predicted by Tifft and a very smooth velocity field,
with a velocity dispersion $12~{\rm Km.s.}^{-1}$.
In Tifft (1977a) \cite{bi:tifft77a} it was claimed that redshift differentials
between pairs of galaxies and between galaxies in clusters take preferred 
values which are various multiples of a basic $72.5 ~{\rm Km.~ s}^{-1}$.   
In Tifft (1977b) \cite{bi:tifft77b} 
the effect was studied for abnormal galaxies.
In Tifft (1978a) \cite{bi:tifft78a} it was claimed that the asymmetry 
in galaxy {\rm H1} profiles can be related directly
to the properties of discrete redshift.   
In Tifft (1978b) \cite{bi:tifft78b} the concept of 
discrete redshift was applied to dwarf {\rm H1} redshifts and line profiles,
and a model of redshift based upon the ultimate discrete levels spaced 
near $12~ {\rm Km}.~ {\rm s}^{-1}$ was developed.   
Most optical redshift data are not accurate to show discreteness directly,  but
it is claimed that 21 cm. data on double galaxies show the effect clearly;    
Tifft (1980) \cite{bi:tifft80} and Tifft (1982a) \cite{bi:tifft82a} 
using the radio data taken by Peterson (1979) \cite{bi:peterson} 
at the {\bf NRAO 300'} telescope has claimed that the effect is
present with a confidence level of $99.9\vb+
Cocke and Tifft (1983) \cite{bi:CT} have claimed
that the effect is present in the {\rm 21 cm.} data on compact groups of 
galaxies taken by Haynes (1981) \cite{bi:haynes} 
and Heleou et al (1982) \cite{bi:HeST} at the Arecibo telescope with a 
confidence level of $99.5\vb+
The optical data of Tifft (1982b) \cite{bi:tifft82b} also shows the 
effect strongly.   The Fisher and Tully (1981) \cite{bi:FT} 
survey of {\rm 21 cm.} redshift was found by 
Tifft and Cocke (1984) \cite{bi:TC} to show sharp periodicities at exact 
multiplies ($\fr{1}{3}$ and $\fr{1}{2}$) of $72.45 ~{\rm Km}.~ {\rm s}^{-1}$;  
the periodicity at $24.1 ~{\rm Km}.~ {\rm s}^{-1}$  
involves galaxies with narrow {\rm 21 cm.} profiles and the $\fr{1}{2}$ 
periodicity at $36.2 ~{\rm Km}.~ {\rm s}^{-1}$ involves galaxies with wide 
profiles,  and there appears to be a progression of periods  
$24\rightarrow 36\rightarrow 72$ for galaxies with higher {\rm 21 cm.} flux
levels as the profile width increases.   Arp and Sulentic (1985) \cite{bi:AS},
using data from the Arecibo telescope of over $100$ galaxies in more than 
$40$ groups found that:  companion galaxies have a higher redshift than 
the dominant galaxy of the cluster,  and that the difference in redshift 
between the dominant and companion galaxies occurs in multiplies of 
$70 ~{\rm Km}.~ {\rm s}^{-1}$.
Sharp (1984) \cite{bi:sharp} suggested that for double galaxies discrete 
redshift might be just a statistical effect.
Newman {\it et al} (1989) \cite{bi:NHT89} suggested that the effect is 
gaussian random noise and that at least one order of magnitude more 
data is needed to confirm the effect.
Crousdale's (1989) \cite{bi:crousdale} study supports Tiffts' work.
Mirzoyan and Vardanyan (1991) \cite{bi:MV} claim that the values of the 
redshifts found preferentially in quasistellar objects essentially coincide 
with the redshifts for which the strong emission lines of Mg II,  C IV,  
Ly $\al$,
in the spectra of these objects fall close to the maximum sensitivity of the
U,  B,  and V light filters.   In this case their effect on the conditions 
for observing quasars is decisive and causes the quasar redshifts to be 
discretized.   Based on a comparison between the observed quasar redshifts 
and the expected values assuming that this explanation is correct,  they 
conclude that the observed effect of quasar redshift discretization is 
caused by observational selection.
Guthrie and Napier (1991) \cite{bi:GN91} confirm the effect in 
near by galaxies,  but whereas Tifft finds 
24.2,  36.3,  or 72.5 Km.$ {\rm s}^{-1}.$ they find
37.2 Km.${\rm s}^{-1}$.
Kruogovenko and Orlov (1992) \cite{bi:KO} find a periodic cycle of 
Seyfert and radio galaxies of about $30{\rm h}^{-1}$Mpc between shells.
Holba {\it et al} (1994) \cite{bi:HHLP} find a non-negligible region where 
two quasar samples and the galaxy sample are simultaneously fairly periodic.
Guthrie and Napier (1996) \cite{bi:GN96} confirm redshift periodicity 
in the local supercluster.
Khodyachikh's (1996) \cite{bi:khod} findings contradict the explanation
of periodicity by selection effects. 
Tifft (1996) \cite{bi:tifft96} finds 72 and 36 Km.${\rm s}^{-1}$ periodicity 
for galaxy samples from the Virgo cluster,  
the Perseus and Cancer supercluster regions,  and local space.
Tifft (1997) \cite{bi:tifft97} studies the redshift of local galaxies for 
quantization and finds that ordinary spiral galaxies with 21 cm. profile 
widths near 200 Km.${\rm s}^{-1}$ show periodic redshifts.
A review of redshift periodicities has been given by Narlikar (1992) 
\cite{bi:nark},  he finds that different data sets of extragalactic objects 
including nearby and distant galaxies and quasars that show statistically 
significant peaks at periodic intervals of redshift;  he says at present the 
data is not complete in any sense but they are substantial enough to make us 
worry about the fundamental assumption that the Universe is homogeneous on a 
large scale.   Moreover,  he claims evidence of this kind has not only 
persisted in spite of rigorous statistical analysis but has grown with time 
so that it cannot be altogether ignored.
Recent conference proceedings covering some aspects of discrete redshift,
Pitucco {\it et al} (1996) \cite{bi:PTDC} 
and Dwari {\it et al} (1996) \cite{bi:DDN}.
Newman {\it et al} (1994) \cite{bi:NHT94} and
Newman and Terzian (1996) \cite{bi:NT} analyses the ``Power Spectrum Analysis''
(PSA) of Yu and Pebbles (1969) \cite{bi:YP},  this is the statistical analysis 
used in discrete redshift studies.
They find that this method generates a sequence of random numbers from 
observational data which,  it was hoped, is exponentially distributed with 
unit mean and unit variance.   The variable derived from this sequence is 
approximately exponential over much of its range but the tail of the 
distribution is far removed from an exponential distribution,  so that
statistical inference and confidence testing based on the tail of the 
distribution is unreliable.   Newman and Terzian go on to say that there 
are six claims of the PSA method which are wrong or involve some hidden
assumptions.   For purposes of illustration consider the first of these.
Let $N$ points $x_j$ be distributed in the interval $0$ to $2\pi$ and let
\be
z_n=N^{-1/2}\sum^N_{j=1}\exp(inx_j).
\lb{eq:NT1}
\ee
Yu and Peebles claim that the ensemble average of $z_n(n\ne 0)$ is
\be
<z_n>=N^{-1/2}\sum<\exp(inx_j)>~~
     =N^{-1/2}\sum\int^{2\pi}_0 \fr{dx_j}{2\pi}\exp(inx_j)=0
\lb{eq:NT2}
\ee
Apparently the correct way to approach this is to suppose that the $x_j$ are
identically distributed and independent deviates with distribution 
${\cal P}(x)$ and to define a characteristic generating function ${\cal F}(n)$ 
by
\be
{\cal F}(n)\equiv~~<\exp(inx)>~~=\int^\infty_{-\infty}\exp(inx)d{\cal P}(x),
\lb{eq:NT3}
\ee
then
\be
<z_n>=N^{1/2}{\cal F}(n).
\lb{eq:NT4}
\ee
If the underlying distribution function ${\cal P}(x)$ is normally distributed 
(gaussian) with a mean $\mu$ and a variance $\si^2$ then
\be
<z_n>=N^{1/2}\exp(-\fr{n^2\si^2}{2}).
\lb{eq:NT5}
\ee
Newman and Terzian also say that astronomers choose too small a frequency 
class interval (bin with) in their frequency histograms,  and that the optimal
is $1+\log_2(N)$,  and this is illustrated by their smoother figurers for the
larger interval.
\subsection{theoretical explanations}
Theories to explain discrete redshift have been devised by Cocke and Tifft
(1983) \cite{bi:CT} and Cocke (1985) \cite{bi:coc85},  
Nieto (1986) \cite{bi:nieto},  and Buitrago (1988) \cite{bi:buit}.   
These theories depend on the introduction of 
a redshift quantum mechanical operator;  
this is essentially equivalent to replacing 
the emitter's four-momentum $P^{a}$ by an operator.
Narlikar and Burbidge (1981) produce an explanation which has a two component
model of the Universe with discrete matter.
An explanation was devised by Barut {\it et al} (1994) \cite{bi:BBNR}
where:  ``It is shown that the energy distribution in this model is
periodic and the periods and density decrease with increasing distance,
in striking agreement with experimental data.''.
Discrete redshift of a quantum energy spectrum in an anisotropic universe 
was found by Lamb {\it et al} (1994) \cite{bi:LCK}.
Discrete redshift might be caused be dislocation solutions to equations,
such solutions have been described by Edelen (1994) \cite{bi:edelen}.
Arp (1996) \cite{bi:arp} argues that there is a ``signal carrier'' for 
inertial mass,  which he calls the machion,  and this gives rise
to periodicity.
Greenberger (1983) \cite{bi:greenberger} and Carvalho (1985) \cite{bi:carv85}
generalizes the quantum commutator to $[q,p]=i\hbar+i\hbar f(q,p)$,  
where $f(q,p)$ is a function,  Carvalho (1997) \cite{bi:carv97} uses this 
to calculate a redshift spectrum with discrete values.
Hill {\it et al} (1990) \cite{bi:HiST} constructed three alternative models 
involving oscillating physics:  i) an oscillating gravitational constant,
this was also studied by Salgado {\it et al} (1996) \cite{bi:SSQ},
ii) oscillating atomic electron mass,  this was ruled out by Sudarsky (1992)
\cite{bi:sudarsky} on the basis of the Braginsky-Panov experiment,  and
iii) oscillating galactic luminosities.   
A varying Hubble parameter has been used by Morikawa (1991) \cite{bi:morikawa} 
as an explanation.
A Voroni cellular model was used by van de Weygaert (1991) \cite{bi:weg}
to explain the Broadhurst {\it et al} result.
Ikeuchi and Turner (1991) \cite{bi:IT} also use a three dimensional Voroni 
tessellation model to explain voids and walls.
Williams {\it et al} (1991) \cite{bi:WPH} find that the Voroni foam model 
predicts a scale length for the galaxy-galaxy correlation function which 
is too large.
Hill {\it et al} (1991) \cite{bi:HST91} suggest a particular coherent 
sinusoidal peculiar velocity field of amplitude $\de/c\approx 3\times 10^3$ 
and wavelength $\lambda\approx 128h^{-1}$Mpc could explain the result of 
Broadhurst {\it et al}.
An oscillating gravitational constant model was constructed by Crittenden and 
Steinhart (1992) \cite{bi:CS} to explain the Broadhurst {\it et al} result.
Dekel {\it et al} \cite{bi:DBPS} argue that the Broadhurst {\it et al} result
suggests a large scale origin for periodicity.
Budinick {\it et al} (1995) \cite{bi:BNRR} built another model to explain 
the Broadhurst {\it et al} result.
The existence of superclusters and voids is perhaps explained by the 
cold dark matter theory of White {\it et al} (1987) \cite{bi:WFDE}.   
Redshift periodicity can be used to probe the correctness of general
relativity Faroni (1997) \cite{bi:faroni}.
Lui and Hu (1998) \cite{bi:LH} suggest an explanation which has the mean free
path of heavy elements absorb system varies regularly with cosmic time.
Farhi {\it et al} (19980 \cite{FGHJ} technique might provide an explanation.
Some papers on the quantum mechanics of large macroscopic
systems are:  Greenberger (1983) \cite{bi:greenberger},  
Agnese (1984) \cite{bi:agnese},  DerSarkissan (1984) \cite{bi:ders84}
and (1985) \cite{bi:ders85},  Carvalho (1985) \cite{bi:carv85},
Silva (1997) \cite{bi:silva}
Capozziello {\it et al} (1998) \cite{bi:CFL},  
and Carneiro (1998) \cite{bi:carneiro}.
\subsection{discrete as opposed to continuous energy spectrum}
It is a common error in non relativistic quantum mechanics to suppose that
quantization implies that there must be a discrete energy spectrum,  see for
example Schiff (1949) \cite{bi:schiff} page 34.
For an infinite potential $V(r)$ there are discrete energy levels;
however for a potential with $E>V$ there are continuous energy levels 
when $E>V$.   Usually however 
a discrete spectrum is usually an indication of a system with quantum rather
than classical properties.   By analogy with non relativistic quantum mechanics
a closed gravitational interacting system would be expected to display 
discrete properties at low energies,  i.e.  when the gravitational field is
weakly interacting,  rather than at high energies when the gravitational field
is strong.   This suggests that cosmology and extra-galactic astronomy - as 
opposed to particle physics,  are the subject areas where a single theory 
combining quantum mechanics and gravity would be necessary.   
There are many unusual dynamical properties of large scale systems,  
such as galaxies,  see for example Roberts (1991) \cite{bi:mdr91} 
and references therein,  and it might be that these 
are directly attributable to quantum corrections to classical theory;  
however here attention is restricted to systems that are characterized by
discrete properties rather than unusual dynamics.
Bohr-Sommerfeld quantization rule have been applied to gravitational systems:
Wereide (1923) \cite{bi:wereide} applied them to spherically symmetric 
spacetimes to find the line element of the electron,  
and Agnese and Testa (1997) \cite{bi:AF} applied them to planetary orbits.
There are discrete approaches to quantum gravity Loll (1998) \cite{bi:loll},
and it might be that discrete structure from near the Planck era is inflated to
present day large scale discrete structure.   Random walk models suggest a
fundamental length of $l\ap10^{35}$cm. Sidharth (1999) \cite{bi:sidharth}.
\subsection{sectional contents}
In section  \ref{sec:RWst} the exact solution for the equation of state 
$p=(\gm-1)\mu$ Robertson-Walker spacetime is derived.   
This is done because the critical parameter $2q_{0}$ 
occurs in latter approximations to the Klein-Gordon equation and these
exact solutions clarify when the occurrence of this critical parameter is an
artifact of the approximations involved;  also the standard approach to 
redshift in Robertson-Walker spacetime is given.   
Section  \ref{sec:remarks} consists of general
remarks on what properties a theory of discrete redshift would be expected to
have and discusses various approaches to constructing a theory.   
Section  \ref{sec:KG}
is devoted to finding approximate solutions to the Klein-Gordon equation in
Robertson-Walker spacetimes.  
Section  \ref{sec:dsvCR} applies these approximate solutions
to theories of discrete redshift where the discreteness of radiation 
originates from the motion of the emitter.   
Section  \ref{sec:dsvKG} discusses how discrete redshift might occur
via the massive Klein-Gordon equation.
Section  \ref{sec:dsvQP} uses the solutions
for the Klein-Gordon equation in the Einstien static universe to induce weak
field metric perturbations of Robertson-Walker spacetime,  these weak metric 
perturbations are used to demonstrate discrete redshift.
\subsection{conventions}
The conventions used are: signature $-+++$, 
early latin indices   $a,b,c\ldots   = 0,1,2,3$,
middle latin indices  $i,j,k\ldots  = 1,2,3$,
Riemann tensor
\be
R^{a}_{.bcd}=2\p_{[c}\Ga^{a}_{d]b}+2\Ga^{a}_{[c|f|}\Ga^{f}_{d]b},
\label{eq:1.1}
\ee
Ricci tensor
\be
R_{bd}=R^{a}_{.bad},
\label{eq:1.2}
\ee
commutation of covariant derivatives
\be
X_{ab;cd}-X_{ab:dc}=X_{ae}R^{e}_{.bcd}+X_{eb}R^{e}_{.acd}.
\label{eq:1.3}
\ee
Relativistic units are not used;  $c$,  $G$,  and $\hb$  
are put explicitly into equation where appropriate,  
this is done in order to clarify the relative
size of the terms in the approximations used,
$c$ is included in the definition of the Hubble constant \ref{eq:hubb}
\section{Robertson-Walker Spacetime.}
\label{sec:RWst}
\subsection{the line element}
The Robertson-Walker line element is
\be
ds^{2}=-c^{2}N(t)^{2}dt^{2}+R(t)^{2}d\Si^{2}_{3},
\label{eq:2.1}
\ee
where $R$ is the scale factor and
\ber
d\Si^{2}_{3}&=& d\ch^{2}+s(\ch)^{2}(d\te^{2}+\sin^{2}\te d\ph^{2}),\nonumber\\
{\rm with}~~~~~~~~~~~~~~~~~~~&&\\
s(\ch)&=&\left\{
\begin{array}{rcl}
\sin(\ch)~~~&{\rm for}&~k=+1\nonumber\\
\ch~~~&{\rm for}&~k=0\\
\sinh(\ch)~~~&{\rm for}&~k=-1.\nonumber
\end{array}
\right.
\label{eq:2.2}
\ear
The lapse function $N$ is arbitrary and depends on the time coordinate used.   
Three common choices for the lapse function are:  i) $N=1$  for which the time
coordinate is the same as the proper time of a co-moving observer,  
ii) $N=R$ 
for which the line element is conformal to the Einstein static Universe,  
and the time coordinate referred to as conformal time,  
and iii) $N= R^{3}$ for which $\Ga^{t}\equiv g^{ab}\vb+{+^{t}_{ab}\vb+}+=0$,  
and the time coordinate referred to as harmonic time.   
For $k=+1$ using the coordinate transformations,  
M\"uller (1939) \cite{bi:muller} 
\ber
\sin(\al)&=&\sin(\te) \sin(\ch),\nonumber\\  
\cos(\bt)&=&\sqrt{1+\cos^{2}(\te) \tan^{2}(\ch)},\nonumber\\       
\cos(\te)&=&\sin(\bt) \sqrt{\sin(\bt)+\tan^{2}(\al)},\nonumber\\
\cos(\ch)&=&\cos(\al) \cos(\bt),
\label{eq:2.3}
\ear
the three-sphere line element becomes
\be
d\Si_{3+}^{2}=d\al^{2}+\cos^{2}(\al) d\bt^{2}+\sin^{2}(\al) d\ph^{2},
\label{eq:2.4}
\ee
and the three-geodesic distance $\om$ takes the simple form
\be
\cos(\om)=\cos(\al)\cos(\al')\cos(\bt-\bt')
          +\sin(\al) \sin(\al')\cos(\ph-\ph ').
\label{eq:2.5}
\ee
\subsection{Taylor series expansion}
The geodesic distance,  or world function is known only for a few special 
cases,  Roberts (1993) \cite{bi:mdr93}.   
For $N=1$,  and for times close to an observers 
time $t_{0}$ the scale factor can be expanded as a Taylor series
\ber
R&=&R_0\left[1+H_{0}\de t-\fr{1}{2}q_{0}H_{0}^{2}\de t^{2}
                 +\fr{1}{6}j_{0}H_{0}^{3}\de t^{3}+O(H_{0}\de t)^{4}\right],\nn
&&{\rm where}\nn
\de t&=&t-t_{0},~~~
\dot{R}\equiv\p_{t} R,\nn  
H&\equiv&\fr{\dot{R}}{R},~~~ 
q\equiv-\fr{R\ddot{R}}{\dot{R}^2},~~~  
j\equiv\fr{\dot{\ddot{R}}R^{2}}{\dot{R}^{3}},
\label{eq:taylor}
\ear
and the the subscripted ``0'' as in $H_0$ refers to the value of the object 
measured by a observer at $t=t_0$.
For arbitrary lapse and $c$ explicit these become
\ber
H&\equiv&\fr{\dot{R}}{cNR},~~~
q\equiv-\fr{1}{\dot{R}^2}\left(\ddot{R}-\fr{\dot{N}\dot{R}}{N}\right),\nn
j&\equiv&\fr{R^2}{\dot{R}^3}\left(\dot{\ddot{R}}-3\fr{\dot{R}\dot{N}}{N}
   -\fr{\dot{R}\dot{N}}{N}+3\fr{\dot{R}\dot{N}^2}{N^2}\right).
\label{eq:hubb}
\ear
\subsection{the field equations}
For the line element \ref{eq:2.1} the Christoffel symbols are
\ber
\{^{t}_{it}\}&=&\{^{i}_{tt}\}=0,\nonumber\\
\{^{t}_{tt}\}&=&\fr{\dot{N}}{N},
~~~\{^{i}_{tj}\}=\de^{i}_{j}\fr{\dot{R}}{R},
~~~\{^{t}_{ij}\}=g_{ij}\fr{\dot{R}}{c^{2}N^{2}R},\nonumber\\
\{^{\al}_{\bt \bt}\}&=&\cos(\al)\sin(\al),
~~~\{^{\al}_{\ph \ph}\}=-\cos(\al)\sin(\al),\nonumber\\
\{^{\bt}_{\bt \al}\}&=&-\tan(\al),
~~~\{^{\ph}_{\ph \al}\}=\cot(\al).
\label{eq:2.8}
\ear
The Riemann tensor is
\ber
R^{(3)}_{ijkl}&=&g^{(3)}_{ik}g^{(3)}_{lj}-g^{(3)}_{li}g^{(3)}_{kj},\nonumber\\
R_{ijkl}&=&\left(\fr{k}{R^2}+\fr{\dot{R}^2}{c^{2}N^{2}R^{2}}\right)
              (g_{ik}g_{lj}-g_{li}g_{kj}),\nonumber\\
R_{titj}&=&\left(-\fr{\ddot{R}}{R}+\fr{\dot{N}\dot{R}}{NR}\right)g_{ij}.
\label{eq:2.9}
\ear
using Einstein's field equations the density $\mu$ and the pressure $p$ for
a perfect fluid are given by the Friedman equation
\be
\fr{8\pi G}{c^{4}}\mu=3\fr{k}{R^{2}}+3\fr{\dot{R}^{2}}{c^{2}N^{2}R^{2}},
\label{eq:2.10}
\ee
and the pressure equation
\be
\fr{8\pi G}{c^{4}}p=-\fr{k}{R^{2}}-\fr{1}{c^{2}N^{2}}
  \left(2\fr{\ddot{R}}{R}+\fr{\dot{R}^{2}}{R^{2}}
                            -2\fr{\dot{N}\dot{R}}{NR}\right).
\label{eq:2.11}
\ee
The conservation equation is
\be
\fr{d}{dR}(\mu R^3)=-3pR^{2}.
\label{eq:2.12}
\ee
compare Weinberg \cite{bi:weinberg} equation 15.1.21.   For the equation 
of state
\be
p=(\gm-1)\mu,
\label{eq:2.13}
\ee
$(\gm-1)$ times the Friedman equation \ref{eq:2.10} minus the pressure 
equation \ref{eq:2.11} is 
\be
0=\fr{3\gm-2}{R^2}\left(k+\fr{\dot{R}^{2}}{c^2N^{2}}\right)
 +\fr{2}{c^2N^{2}}\left(\fr{\ddot{R}}{R}-\fr{\dot{N}\dot{R}}{NR}\right),
\label{eq:2.14}
\ee
evaluating this at $t=t_{0}$ and using \ref{eq:hubb}
\be
0=k(3\gm-2)+H_{0}^{2}R_{0}^{2}(3\gm-2-2q_o),
\label{eq:2.15}
\ee
which for given $\gm$ and $q$ determines the sign of $k$ and hence whether 
the Robertson-Walker line-element is open or closed,
in the particular case $\gm=1$ and $N=1$, \ref{eq:2.15} reduces to 
Weinberg \cite{bi:weinberg} equation 15.2.5.   For the equation 
of state \ref{eq:2.13} the conservation equation \ref{eq:2.12} is
\be
\gm\mu^{1-1/\gm}\fr{d}{dR}\left(\mu^{\fr{1}{\gm}}R^{3}\right)
=\mu_{,R}R^{3}+3\gm\mu R^{2}=0.
\label{eq:2.16}
\ee
Integrating
\be
\mu^{\fr{1}{\gm}}R^{3}=a,
\label{eq:2.17}
\ee
where $a$ is a constant.   Thus $\mu$ is proportional to $R^{-3\gm}$ so that
\be
\fr{\mu}{\mu_{0}}=\left(\fr{R_{0}}{R}\right)^{3\gm},
\label{eq:2.18}
\ee
$\mu_{0}$ is also given by the Friedman equation \ref{eq:2.10} at $t_{0}$,  
combining with \ref{eq:2.18} this gives
\be
\mu=\fr{3c^{4}}{8\pi G}\left(H_0^2+\fr{k}{R_0^2}\right)
     \left(\fr{R_0}{R}\right)^{3\gm}.
\label{eq:2.19}
\ee
where a factor of $c$ is included in our definition of $H$ \ref{eq:hubb}.
Substituting \ref{eq:2.19} into \ref{eq:2.10} the Friedman equation becomes
\be
\left(\fr{\dot{R}}{cN}\right)^2+k=\left(\fr{R}{R_{0}}\right)^{2-3\gm}
      \left(k+H_0^2R_0^2\right).
\label{eq:2.20}
\ee
\subsection{the general solutions for $\gamma$-equation of state}
For $3\gm=2$ the solutions of the Friedman equation is the generalized Milne 
universe
\be
N=1,~~~R=R_{0}(1+H_{0}t),
\label{eq:2.21}
\ee
for which the scale factor $R$ is just given by \ref{eq:taylor} up to first 
order.   For $3\gm\ne2$,  define
\be
R_{\rh}\equiv R_{0}\left(H_0^2R_0^2+k\right)^{\fr{1}{3\gm-2}},
\label{eq:2.22}
\ee
then equation \ref{eq:2.20} fixes the constant in the solution of 
Vajk (1967) \cite{bi:vajk}
\ber
k&=&1,~~~cN=R=R_{\rh}\left[\sin(\fr{3\gm-2}{2}\et)\right]
                                        ^{\fr{2}{3\gm-2}},\\
k&=&0,~~~~N=1,
      ~~~R=R_{0}\left(\fr{3\gm cH_{0}t}{2}\right)^{\fr{2}{3\gm}},\\
k&=&-1,~~cN=R=R_{\rh}\left[\sinh(\fr{3\gm-2}{2}\et)\right]^{\fr{2}{3\gm-2}},
\label{eq:2.23-25}
\ear
where $\et$ is the time coordinate used when $N=R$.   These solutions can be 
expanded around the origin $t=0$ of the time coordinate
\be
R=R_{\rh}\left(\fr{3\gm-2}{2}\right)^{\fr{2}{3\gm-2}}
     \left[1-\fr{k}{6}\left(\fr{3\gm-2}{2}\right)\et\right]^{2}
    +O\left(\fr{3\gm-2}{2}\et\right)^{4}.
\label{eq:2.26}
\ee
Using proper time,  for $\gm=\fr{4}{3}$
\be
R=R_{0}\left[1+kH_0^2R_0^2
         -\fr{kc^{2}t^{2}}{R_{0}^{2}}\right]^{\fr{1}{2}},
\label{eq:2.27}
\ee
and for $\gm=1$ there is the expansion round the origin
\be
R=R_{\rh}\left[1-\fr{k}{4}\left(\fr{ct}{R_{\rh}}\right)^{2}
           -\fr{k}{48}\left(\fr{ct}{R_{\rh}}\right)^{4}  
           +O\left(\fr{ct}{R_{\rh}}\right)^{6}\right],
\label{eq:2.28}
\ee
and also an expansion in powers of $t^{\fr{2}{3}}$.   
The substitution $t\rightarrow t-t_{0}$ 
merely shifts the origin of the time coordinate,  
i.e. the singularity of the spacetime moves from $t=0$ to $t=t_{0}$.   
Replacing $t$ by
\be
t=t_{0}+\de t,~~~\de t=t-t_{0},
\label{eq:2.29}
\ee
in all of the above gives back the Taylor expansion \ref{eq:taylor}.
\subsection{the geodesics}
The Lagrangian for geodesics is
\be
2{\cal L}=g_{ab}\fr{dx^{a}}{ds}\fr{dx^{b}}{ds},
\label{eq:2.30}
\ee
for timelike,  null,  and spacelike geodesics $2{\cal L}=-1,~0,~+1$ 
respectively.
A co-moving geodesic is a timelike geodesic with $\fr{dx^{i}}{ds}=0$,   
the properties and \ref{eq:2.30} imply that the co-moving velocity vector 
is always,  irrespective of the geometry of the spacetime,  given by
\be
U^{a}=\pm\fr{dx^{a}}{ds}=\pm\left(\fr{1}{cN},0\right).
\label{eq:2.31}
\ee
For null geodesics in Robertson-Walker spacetime
\be
0=-c^{2}N^{2}\left(\fr{dt}{d\Om}\right)^{2}
+g_{ij}\fr{dx^{i}}{ds}\fr{dx^{j}}{ds}.
\label{eq:2.32}
\ee
using the three-sphere distance \ref{eq:2.5},  
\ref{eq:2.32}  can be solved to give
\be
\Om=\int\fr{cN}{R}dt+\om,
\label{2.33}
\ee
thus the radiation vector is
\be
k_{a}=\Om,a =\left(\fr{cN}{R},\om_{i}\right).
\label{eq:2.34}
\ee
The redshift is given by the equations,  Ellis (1971) \cite{bi:ellis}
\be
1+z=\fr{ds_{0}}{ds}
   =\fr{\lm_{0}}{\lm}
   =\fr{\nu}{\nu_{0}}
   =\fr{U^{a}k_{a}}{(U^{a}k_{a})_{0}},
\label{eq:2.35}
\ee
as in \ref{eq:2.5} the subscript $"0"$ refers to the observer.   
The quantities at the emitter have no subscript.  
$\nu$ and $\lm$ are the frequency and wavelength 
of the radiation;  $s$ is the proper time.   
Equations \ref{eq:2.31},  \ref{eq:2.34},  and \ref{eq:2.35} 
give independently of the choice of $N$,
\be
1+z=\fr{R_{0}}{R}.
\label{eq:2.36} 
\ee
For the Taylor series expansion \ref{eq:2.5} this becomes
\ber
z=&-&H_0(t-t_{0})
  +\left(1+\fr{1}{2}q_0\right)H_0^{2}(t-t_{0})^{2}\nn
  &-&\left(1+q+\fr{1}{6}j_{0}\right)H_{0}^{3}(t-t_{0})^{3}
  +O\left(H_{0}(t-t_{0})\right),
\label{eq:2.37}
\ear
the observed values of $H_{0}$ and $q_{0}$ are 
$H_{0}=2.4\pm0.8.10^{-18}~{\rm sec}.^{-1}$ and $q_{0}=1\pm1$,
for a recent review see Freedman (1999) \cite{bi:freedman}.
$z$ is found to take the discrete values
\be
z_{n}=\fr{nv_{I}}{c},
\label{eq:2.38}
\ee
where $n$ is an integer and $v_{I}=72.2\pm0.2~{\rm Km}.~{\rm sec}.^{-1}$,  
or perhaps a sixth  of this value.
\section{Remarks on Explanations of Discrete Redshift.}
\label{sec:remarks}
\subsection{three distinctions categorizing discrete redshift}
There are {\sc three} distinctions that should be made to categorize any 
theory of discrete redshift.   The {\sc first} is whether the discrete 
properties enter via the radiation connecting the emitter and observer,  
or by the motion of either (or both) the emitter and observer.   
The {\sc second} is whether the emitter has real discrete motion,  
or only apparent discrete motion,  or neither.   
The {\sc third} is whether the effect is due to quantizing the whole
system,  or part of it,  or is not due to quantum mechanics at all.
\subsection{via connecting radiation}
To elaborate on the {\sc first} distinction,   from equation \ref{eq:2.35} 
it is apparent that discrete redshift must come from discrete differences in 
the ratio of the observer's and emitters proper time 
$\fr{ds_{obs}}{ds_{emm}}$.   
The underlying structure of spacetime may be so unusual as to forbid the
introduction of a time-like four-vector $U^{a}$ 
or a null tangent vector $k_{a}$;
however if this can be done then there are two choices for the origin of 
discrete properties:  either the particles time-like four-vector or the 
connecting radiation appears to take discrete values.   If the metric itself
takes discrete values then both of these would presumably occur.
\subsection{real discrete motion}
To elaborate on the {\sc second} distinction.   {\it Real discrete motion} 
means that the emitter's motion is discrete no matter how it is measured.   
{\it Apparent discrete motion} means that the motion of the emitter merely 
appears to be discrete to the observer;  
to put this another way if the observer chooses to
observe from a different vantage point the observer might not measure the
same discrete motion of the emitter.   Real discrete motion has the 
disadvantage that there would have to be boundaries at the edges of where the
emitter's four-velocity jumps,  these boundaries would lead to other effects,
such as refraction and reflection on the boundary  surface.   Such effects
have not been observed in association with measurements of discrete redshift.
Monet and Deharveny (1977) \cite{bi:MD} do not observe what would be expected 
from real discrete motion,  see subsection \ref{sec:obs}.
The existence of superclusters and voids,   
Sanders et al (1991) \cite{bi:sau},  however
might be an example of an effect of real discrete motion.
The quantum perturbations of Grishchuk (1997) \cite{grishchuk},
Yamammoto {\it et al} (1996) \cite{YST},  and 
Modanese (2000) \cite{modanese}
might give real discrete motion.
\subsection{whole system quantization}
To elaborate on the {\sc third} distinction.   The idea of the whole system  
being treated quantum mechanically is essentially that of quantum cosmology.   
If part of the system is quantized it could be either the connecting radiation 
or the emitting matter.   Of course there is the possibility that discrete 
redshift may not have a quantum origin at all,  for example,  it might be of 
fluid dynamical origin.
\subsection{quantum emitter}
In this paper the idea that the discrete properties originate in the
quantum treatment of the emitting matter  will be pursued.   Before proceeding
with this some remarks will be made on the possibility of discrete redshift 
originating in the properties of the connecting radiation from quantum
cosmology.   There are several possibilities which might give discrete 
connecting radiation,  here two will be mentioned.   
The {\sc first} is that the connecting radiation could 
undergo a scattering process which makes it discrete,   
the scattering process would have to occur in a wide variety 
of circumstances in order to explain the many scales and wavelengths 
over which discrete redshift occurs.   
The {\sc second} is that solutions 
to Maxwell's equations in Robertson-Walker spacetime involve discrete 
properties from spherical harmonics.   
This will be discussed further in Section 5.
\subsection{quantum cosmology}
Quantum cosmology consists of a large amount of theory in which the whole
Universe is considered quantum mechanically 
(see Tipler (1986) \cite{bi:tipler} and Kiefer (1999) \cite{bi:kiefer} 
for reviews);
and it is the obvious framework in which to start looking for an explanation 
of discrete redshift.   There might be a quantum analog of the Friedman 
equation \ref{eq:2.10} which possess solutions in which the metric is discrete.
Developing quantum cosmology as it is found in the literature produced nothing
along these lines,  therefore four more simplistic approaches where attempted:
as it would be anticipated that even a simple model of a quantum expanding
gas would produce a qualitative result which involved discrete redshift.
The dynamics of Robertson-Walker spacetimes are determined by the Friedman
equation \ref{eq:2.10} and the conservation equation \ref{eq:2.12};   
the {\sc first} approach consisted of applying naive operator substitutions 
to the Friedman equation;
by this is meant replacing each symbol in the equation by an operator of the 
form $-ia\p_{b}$ where $a$ is chosen to be the most general combination of 
$\hb$,  $c$,  $G$,  $R$,  $R_{0}$ 
which provides a dimensionally correct substitution for the symbol,
and $b$ can be variously considered to be a four-vector index or a time index
etc\ldots.   
This approach failed because it produced inconsistencies when applied
to the conservation equation and it was impossible to eliminate $R$.   
A method similar to this which works has been constructed by 
Rosen (1993) \cite{bi:rosen},  see also 
Capozziello {\it et al} (1998) \cite{bi:CFL}.
The {\sc second} approach was to apply n\'aive operator substitutions 
to the Newtonian analog of the Friedman equation 
(see for example Weinberg (1972) \cite{bi:weinberg}).   
It might be anticipated that this would produce a well-defined theory 
because it involves only a quantum generalization of Newtonian cosmology,  
however in common with much of Newtonian cosmology $\dot{R}$ and $c$ 
occur in places that lead to inconsistencies.   
The {\sc third} approach is to note that there are special
relativistic gravitational theories,   some of which are discussed in 
North (1965) \cite{bi:north}.   
It would be hoped that they would lead to equations with $\dot{R}$ and $c$ 
in consistent places,  however this approach also failed.   
The {\sc fourth} approach is that Schr\"odinger's p.63 (1956) 
\cite{bi:schro56} derivation of Robertson-Walker redshift by 
thermodynamic analogy might be extendable to include quantum mechanics
and thence discrete redshift.
\subsection{constituent quantization}
Assuming that the Robertson-Walker spacetime remains correct,  it is 
necessary to identify what constituents of its contents needs to be quantized
and by what mechanism.   In general relativity the co-moving emitter travels
on geodesics for which
\be
2 {\cal L}=U_{a}U^{a}=\fr{1}{m}p_{a}p^{a}.
\label{eq:3.1}
\ee
This equation can be n\'aively quantized by using the operator substitution
\be
p_{a}\rightarrow i\hb \na_{a},
\label{eq:3.2}
\ee
giving the Klein-Gordon equation
\be
\left(\Box-\fr{m^{2}c^{2}}{\hb^{2}}\right)\psi=0,
\label{eq:3.3}
\ee
but it is not clear what the interpretation of 
\ref{eq:3.3} is in the present context.
For example,  should the mass in the Klein-Gordon equation be interpreted as
the mass of the emitting galaxy or the emitting atom or something in between,
and has the single particle theory described by 
\ref{eq:3.1} become a many particle theory described by \ref{eq:3.3}?   
Here what the Klein-Gordon field describes will be discussed later.   
The Universe will be taken to have closed Robertson-Walker geometry,  
because it is for closed systems that discrete properties usually occur.
\subsection{references for the KG equation in RW spacetime}
The Klein-Gordon equation in Robertson-Walker spacetimes has been studied 
for a variety of purposes by:  
Schr\"odinger (1939) \cite{bi:schro39},  (1956) \cite{bi:schro56},  
M\"uller (1940) \cite{bi:muller},  
Lifshitz (1946) \cite{bi:lif46},  
Lifshitz and Khalatnikov (1963) \cite{bi:LK},  
Ford (1976) \cite{bi:ford},  
Barrow and Matzner (1980) \cite{bi:BM},  
and Klainerman and Sarnak (1981) \cite{bi:KS}.   
Maxwell's equation in Robertson-Walker spacetimes has been studied by 
Schr\"odinger (1940) \cite{bi:schro40},  and Mashhoon (1973) \cite{bi:mash}.
Dirac's equation in Robertson-Walker spacetime has been 
studied by Schr\"odinger (1938) \cite{bi:schro38},  (1940) \cite{bi:schro40},  
and Barut and Duru (1987) \cite{bi:BD}.
\section{The Klein-Gordon Equation in Robertson-Walker Spacetime.}
\label{sec:KG}
\subsection{spherical harmonics}
In Robertson-Walker spacetime the Klein-Gordon equation is
\be
0=-\fr{1}{NR^3}\left(\fr{R^3\dot{\ph}}{N}\right)^\bullet
  +\fr{c^2}{R^2}K(\ph)-\fr{c^4m^2}{\hb}\phi,
\label{eq:4.1}
\ee
In the coordinates \ref{eq:2.4},  $K(\ph)$ takes the form
\be
K(\ph)=\sec(\al){\rm cosec}(\al)
           \left(\cos(\al)\sin(\al) \ph_{\al}\right)_{\al}    
      +\sec^{2}(\al) \ph_{\bt \bt}+{\rm cosec}^{2}(\bt) \ph_{\al \al}.
\label{eq:4.2}
\ee
Define 
\be
Y\equiv \sin^{|n|}\al)\cos^{|p|}(\al)\exp~i(|n|\ph+|p|\bt),
\label{eq:4.3}
\ee
then
\be
\fr{K(\ph)}{Y}=-\left(|n|+|p|\right)\left(|n|+|p|+2\right).
\label{eq:4.4}
\ee
The coordinate ranges $0<\al<\fr{1}{2}\al$ and $0<\bt,  \ph<2\pi$   
imply that $n$ and $p$ are integers,  thus defining
\be
l\equiv|n|+|p|,
\label{eq:4.5}
\ee
gives
\be
K(\ph)=Y^{i}_{.~|i}=-l(l+2)Y,
\label{eq:4.6}
\ee
and 
\be
Y_{i}Y^{i}_{.}=-l^{2}\fr{Y^{2}}{R^{2}}.
\label{eq:4.7}
\ee
The complex conjugate to equation \ref{eq:4.6} and \ref{eq:4.7} also holds,  
furthermore
\be
Y_{i}Y^{\dag i}{.}=\left[n^{2}(\cot^{2}(\al)+{\rm cosec}^{2}(\al)) 
                  +p^{2}(\tan^{2}(\al)+\sec^{2}(\al))-2np\right]Y~Y^{\dag},
\label{eq:4.8}
\ee
and 
\be
(Y~Y^{\dag})^{i}_{.|i}=4\sin^{|2n|}(\al)\cos^{|2p|}(\al)
                   \left(n^{2}\cot^{2}(\al)+p^{2}\tan^{2}(\al)-2np-p-n\right),
\label{eq:4.9}
\ee
also
\ber
Y~Y_{i|j}&=&Y_{i}Y_{j},~~~~~~  i\ne j,
\label{eq:4.10}
\\
i(Y~Y^{\dag}_{\al}-Y^{\dag}Y_{\al})&=&0,\nonumber\\
i(Y~Y^{\dag}_{\bt}-Y^{\dag}Y_{\bt})&=&2pY~Y^{\dag},\nonumber\\
i(Y~Y^{\dag}_{\bt}-Y^{\dag}Y_{\ph})&=&2nY~Y^{\dag},
\label{eq:4.11}
\ear
where the dagger $"\dag"$ denotes complex conjugate.   
Equations \ref{eq:4.8},  \ref{eq:4.9},  \ref{eq:4.10},  
and \ref{eq:4.11} cause difficulties when considering the stress of a 
Klein-Gordon field.   Defining the dimensionless scalar field 
\be
\ph Y=\left(\fr{R}{R_{0}}\right)^{-\fr{3}{2}}
      \left(\fr{N}{N_{0}}\right)^{\fr{1}{2}}\psi,
\label{eq:4.12}
\ee
the Klein-Gordon equation \ref{eq:4.1} for $\ps\ne0$ becomes
\be
0=\fr{\ddot{\ps}}{\ps}+X+c^{2}l(l+2)\fr{N^{2}}{R^{2}}
 +\fr{c^{4}m^{2}N^{2}}{\hb},
\label{eq:4.13}
\ee
where
\be
X\equiv\fr{1}{2}\left(\fr{\ddot{N}}{N}-3\fr{\ddot{R}}{R}\right)
       -\fr{3}{4}\left(\fr{\dot{R}}{R}-\fr{\dot{N}}{N}\right).
\label{eq:4.14}
\ee
\subsection{Hill's equation}
For $R$ and $N$ consisting of trigonometric functions this equation is 
similar to Hill's equation,  see for example p.406 Whittaker and Watson 
(1927) \cite{bi:WW}.   For the Einstein static universe there is the solution
\be
\ph=C_{+}\exp(i\nu t)+C_{-}\exp(-i\nu t),~~~~~~~~~
\nu^{2}=\fr{c^{2}l(l+2)}{R_{0}^{2}}+\fr{m^{2}c^{4}}{\hb^{2}},
\label{eq:4.15}
\ee
where $C_{+}$ and $C_{-}$ are constants.   
For the generalized Milne universe there is the massless solution,  
Schr\"{o}dinger (1939) \cite{bi:schro39},
\ber
\ph &=&C_{+}\ta^{\fr{1}{2}(1+rt)}+C_{-}\ta^{\fr{1}{2}(1-rt)},\nn
{\rm with}~~~~~~~~~~\ta&=&-(2H_{0}R_{0}R^{2})^{-1},\\
{\rm and}~~~~~~~~~~rt^{2}&=&1-\fr{c^{2}l(l+2)}{H_{0}^{2}R_{0}^{2}}.\nonumber
\label{eq:4.16}                 
\ear
For the closed $\gm=\fr{4}{3}$ 
spacetime \ref{eq:2.18} there is the massless solution,  
Lifshitz (1946) \cite{bi:lif46},
\be
\ph={\rm cosec}(\et)\left(C_{+}\exp(+i(l+1)\et)+C_{-}\exp(-i(l+1)\et)\right).
\label{eq:4.17}
\ee
In general \ref{eq:4.13} is intractable and it is necessary to use approximate 
WKB solutions to it.
\subsection{WKB approximation}
The WKB approximation is derived as follows,   
Alvarez (1989) \cite{bi:alvarez}.   
Assume the differential equation can be put in the form
\be
y''+\fr{f(x)y}{\hb^2}=0,
\label{eq:4.18}
\ee
where $y'=\fr{dy}{dx}$.   Let
\be
y=\exp(\fr{iz}{\hb}),
\label{eq:4.19}
\ee
then \ref{eq:4.18} becomes
\be
i\hb z''-z'^{2}+f(x)=0,
\label{eq:4.20}
\ee
taking $\hb$ to be small
\be
z'=\pm \sqrt{f},
\label{eq:4.21}
\ee
which gives the first order approximation
\be
y=\exp\left(\fr{i}{\hb}\int\sqrt{f}dx\right).
\label{eq:4.22}
\ee
substituting the derivative of \ref{eq:4.21},  $z''$,  into \ref{eq:4.20}
\be
z'^{2}=f\left(1\pm\fr{i\hb f'}{2f^{\fr{3}{2}}}\right),      
\label{eq:4.23}
\ee
taking the square root and disregarding terms in $\hb^2$,
\be
z'=\pm\sqrt{f}+\fr{i\hb}{4}\fr{f'}{f},
\label{eq:4.24}
\ee
where all combinations of sign are possible,  
choosing the sign in front of the second term to be positive,
using equation \ref{eq:4.10},
and integrating
\be
y=f^{-\fr{1}{4}}\exp\left(\pm\fr{i}{\hb}\int\sqrt{f}dx\right).
\label{eq:4.25}
\ee
The second order approximation is then given by the linear combination
\be
y=f^{-\fr{1}{4}}\left\{C_{+}\exp(+\fr{i}{\hb}\int\sqrt{f}dx)
                    +C_{-}\exp(-\fr{i}{\hb}\int\sqrt{f}dx)\right\}.
\label{eq:4.26}
\ee
Substituting back into \ref{eq:4.18} this approximation holds if
\be
\fr{f}{\hb^{2}}>\fr{5}{16}\fr{f''}{f^{2}}.
\label{eq:4.27}
\ee
\subsection{WKB applied to the KG equation}
Applying the WKB approximation to the Klein-Gordon equation 
\ref{eq:4.13} implies that $\ph$ is of the form
\be
\ph=A(t)Y(\al,\bt,\gm)\left\{C_+\exp(+i\bar{\nu}(t))
                           +C_-\exp(-i\bar{\nu}(t))\right\},
\label{eq:4.28}
\ee
where $C_{+}$ and $C_{-}$ are dimensionless constants.  
There are two functions to be determined.   
The {\sc first} is the dimensionless frequency $\nu $,
\be
\nu .t=\bar{\nu}=\int\left(X+c^{2}l(l+2)\fr{N^{2}}{R^{2}}
                            +\fr{c^{4}m^{2}N^{2}}{\hb}\right)^{\fr{1}{2}}dt,
\label{eq:4.29}
\ee
where the constant of integration is a phase factor which 
is taken to vanish here.   When $t$ is proper time $\nu$ 
is referred to as the proper frequency.
The {\sc second} is the dimensionless amplitude
\ber
A&=&\left(\fr{R}{R_{0}}\right)^{-\fr{3}{2}}
   \left(\fr{N}{N_{0}}\right)^{\fr{1}{2}}A_\ps\nonumber\\
 &=&D\left(\fr{R}{R_{0}}\right)^{-\fr{3}{2}}
   \left(\fr{N}{N_{0}}\right)^{\fr{1}{2}}  
   \left(X+c^{2}l(l+2)\fr{N^{2}}{R^{2}}
         +\fr{c^{4}m^{2}N}{\hb^{2}}\right)^{-\fr{1}{4}},
\label{eq:4.30}
\ear
where $D$ is a dimensional constant added in order to keep $A$ dimensionless.
Both the frequency $\nu$ and the amplitude $A$ are sensitive to the choice of 
time coordinate;  for example in proper,  conformal,  and harmonic times 
respectively
\ber
N=1&:&X=-\fr{3}{4}\left(2\fr{\ddot{R}}{R}+\fr{\dot{R}^{2}}{R^{2}}\right)
       =\fr{3}{4}\left(\fr{8\pi Gp}{c^{4}}+\fr{c^{2}}{R^{2}}\right),\nonumber\\
N=R&:&X=-\fr{R_{,\et \et}}{R},\nonumber\\
N=R^3&:&X=0.
\label{eq:4.31}
\ear
The amplitude $A$ is computed from straightforward substitution;
however the frequency integral \ref{eq:4.29} usually has to be approximated.
\subsection{the Taylor series approximation for the frequency}
For the Taylor series approximation \ref{eq:taylor}
\be
X=\fr{3}{4}\left\{(2q_{0}-1)H_{0}^{2}
                +2(j_{0}-1)H_{0}^{3}(t-t_{0}) 
                +O(t- t_{0})^{2}\right\}.
\label{eq:4.32}
\ee
Up to second order in the Taylor series expansion \ref{eq:taylor} this 
expression is time dependent,  also it depends on the quantity $(2q_{0}-1)$ 
from \ref{eq:2.15} this is just the critical number which determines whether a 
pressure free $\gm=1$ spacetime is open or closed.   
Thus any prediction of $(2q_{0}-1)$ using \ref{eq:4.32} is just an artifact 
of the approximations involved rather than a prediction based on $\gm=1$ 
spacetime,  this is the reason that it is necessary to work with exact 
solutions to Einstein's equations rather than use Taylor series approximations.
From \ref{eq:4.32} and \ref{eq:4.29}
\ber
\nu&=&\left\{\fr{3}{4}(2q_{0}-1)H_{0}^{2}
             +\fr{c^{2}l(l+2)}{R_{0}^{2}}
                 +\fr{c^{4}m^{2}}{\hb^{2}}\right\}^{\fr{1}{2}}\nonumber\\
   &+&\fr{1}{4}\left\{\fr{3}{4}(2q_{0}-1)H_{0}^{2}
                  +\fr{c^{2}l(l+2)}{R_{0}^{2}}
                  +\fr{c^{4}m^{2}}{\hb^{2}}\right\}^{-\fr{1}{2}}
   .\left\{2(j_{0}-1)H_{0}^{2}
              -\fr{2c^{2}l(l+2)}{R_{0}^{2}}\right\}H_{0}(t-t_{0})\nonumber\\
   &+&O(t-t_{0})^{2}.
\label{eq:4.33}
\ear
\subsection{the dimensionless frequency for the $\gamma$-solutions}    
For perfect fluids with $N=R$ the solution to Einstein's field equations 
with $k=1$ gives
\be
X=\fr{3\gm-2}{2}+\fr{3\gm-4}{2}\cot^{2}(\fr{3\gm-2}{2}\et).
\label{eq:4.34}
\ee
For $m=0$ the frequency $\bar{\nu}$ is
\be
\bar{\nu}=\int\lf\{\fr{3\gm-2}{2}
                    +\fr{3\gm-4}{2}\cot^{2}(\fr{3\gm-2}{2}\et)
                    +l(l+2)\rt\}^{\fr{1}{2}}d\et.
\label{eq:4.35}
\ee
Assuming that $l$ is large and expanding
\be
\bar{\nu}=\sqrt{l(l+2)}\lf\{\et+\fr{1}{2l(l+2)}
     \left(\et+\fr{3\gm-4}{2-3\gm}\cot(\fr{3\gm-2}{2}\et)\right)\rt\}
          +O(l^{-4}).
\label{eq:4.36}
\ee
For $m\ne 0$ it is assumed that $\hb$ is small in line with the assumption 
needed for the {\rm WKB} approximation.  
Equivalently,  expressions for the wavelength are of the form
\be
\fr{1}{\lm_{total}^{2}}=\fr{1}{\lm_{compton}^{2}}\pm\fr{1}{\lm_{geometry}^{2}},
\label{eq:4.37}
\ee
where $\lm_{compton}=\fr{\hb}{mc}$ is the Compton wavelength.   
The assumption that $\hb$ is small implies that 
the total wavelength $\lm_{total}$ is dominated by the Compton term. 
Usually,  as can be seen from \ref{eq:4.33} 
the sign in front of the geometric contribution is positive,  
thus $\lm_{total}$ is shorter than the Compton wavelength.
Expanding in $\hb$ the frequency is
\ber
\bar{\nu}&=&\int\left\{\fr{3\gm-2}{2}
                   +\fr{3\gm-2}{2}\cot^{2}(\fr{3\gm-2}{2}\et)
                   +l(l+2)
               +\fr{m^{2}c^2R^{2}}{\hb^{2}}\right\}^{\fr{1}{2}}d\et\nonumber\\
&=&\fr{mc}{\hb}I_{1}+\fr{\hb}{2mc}\left\{\fr{3\gm-2}{2}  
                   +l(l+2)I_{2}  
                   +\fr{3\gm-4}{2}I_{3}\right\}+O(\hb^{3}),
\label{eq:4.38}
\ear
where
\be
I_{1}\equiv\int R~d\et,~~~
I_{2}\equiv\int\fr{d\et}{R},~~~
I_{3}\equiv\int \cot^{2}(\fr{3\gm-2}{2}\et)\fr{d\et}{R}.
\label{eq:4.39}
\ee
Evaluating these integrals for $\gm=\fr{4}{3}$ and $\gm=1$ gives
\ber
\bar{\nu}&=&\fr{cmR_{0}}{\hb}
\left(H_0^2R_0^2+1\right)^{\fr{1}{2}}\cos(\et)\nonumber\\
         &+&\fr{\hb(l+1)^{2}}{2cmR_{0}}
 \left(H_0^2R_0^2+1\right)^{-\fr{1}{2}}
                    \ln|\tan\fr{\et}{2}|\nonumber\\
         &+&O(\hb^{3}),
\label{eq:4.40}
\ear
and
\ber
\bar{\nu}&=&\fr{cmR_{0}}{\hb}\left(H_0^2R_0^2+1\right)
      \left\{\fr{\et}{2}-\cos(\fr{\et}{2})\sin(\fr{\et}{2})\right\}\nonumber\\
         &+&\fr{\hb}{2cmR_{0}}\left(H_0^2R_0^2+1\right)
 \cot(\et)\left\{\fr{1}{3}\cot^{2}(\fr{\et}{2})
                   -\fr{1}{2}(l^{2}+2l+\fr{1}{2})\right\}\nonumber\\
         &+&O(\hb^{3}),
\label{eq:4.41}
\ear
respectively.
\subsection{further approximation}
Using proper time \ref{eq:2.22} and \ref{eq:4.31} 
give for the expansion around the origin
\be
X=\fr{3(3\gm-2)}{4R_{\rh}^{2}}+O(t).
\label{eq:4.42}
\ee
Defining
\be
l'^{2}\equiv l^{2}+2l+\fr{3(3\et-2)}{4},
\label{eq:4.43}
\ee
the proper frequency is
\be
\nu=\nu_{\rh}+O(t),
\label{eq:4.44}
\ee
where
\be
\nu_{\rh}^{2}=\fr{c^{2}l'^{2}}{R_{\rh}^{2}}+\fr{c^4m^2}{\hb^{2}}.
\label{eq:4.45}
\ee
For $\gm=1$,  using \ref{eq:4.31} and \ref{eq:2.28} 
the integral for the proper frequency is
\be
(\bar{\nu}_{,t})^{2}=\nu_{\rh}^{2}
       +\fr{c^{2}l'^{2}}{2R_{\rh}^{2}}\left(\fr{ct}{R_{\rh}}\right)^{2}
       +\fr{11c^{2}l'^{2}}{48R_{\rh}^{2}}\left(\fr{ct}{R_{\rh}}\right)^{4}
       +O\left(\fr{ct}{R_{\rh}^{2}}\right)^6,
\label{eq:4.46}
\ee
expanding the square root and integrating gives the proper frequency
\be
\nu=\nu_{\rh}\left\{1 
+\fr{c^{2}l'^{2}}{12\nu_{\rh}R_{\rh}^{2}}\left(\fr{ct}{R_{\rh}}\right)^{2}
+\left(\fr{11}{3}\nu_{\rh}-\fr{c^{2}l'^{2}}{R_{\rh}^{2}}\right)
 \fr{c^{2}l'^{2}}{160\nu_{\rh}^{2}R_{\rh}^{2}}\left(\fr{ct}{R_{\rh}}\right)^{4}
+O\left(\fr{ct}{R_{\rh}}\right)^{6}\right\}.
\label{eq:4.47}
\ee
\section{Discrete Redshift via the Connecting Radiation.}
\label{sec:dsvCR}
\subsection{explanation using the frequency of the connecting radiation}
The simplest way to produce a theory of discrete redshift is to note
that in \ref{eq:2.35}
$z$ has an expansion in terms of the frequency $\nu$ 
and that the solutions to the massive Klein-Gordon equation also involve 
a frequency.   From \ref{eq:4.31} this frequency depends on the choice 
of time coordinate,  however the $l$ dependent term is usually larger 
than the $X$ term so that this choice only makes a small difference;  
because of this it is sufficient to use the proper frequency.   
Choosing the massless proper frequency \ref{eq:4.44},
the equation for the redshift \ref{eq:2.35} becomes
\be
\nu_{0}(1+z)=\fr{cl'}{R_{\rh}}+O(t),
\label{eq:5.1}
\ee
which implies that
\be
l'\ap(1+z)\fr{\nu_0}{H_0}\left(\fr{3\gm-2}{2q_0+2-3\gm}\right)^\fr{1}{2}
     \left(\fr{2q_0}{2q_0+2-3\gm}\right)^{\fr{1}{3\gm-2}},
\label{eq:5.2}
\ee
as $\fr{\nu_0}{H_0}\sim 10^{26}-10^{32}$,  $l$ must be a very large number,  
as noticed by Schr\"odinger (1939) \cite{bi:schro39}.
\subsection{preservation of proper frequency}
Equation \ref{eq:5.1} depends on $z, 1,  \nu_0, c, H_0,  q_0$ and $\gm$.   
$z$ varies and  $\nu_o,  H_0$ and $q_0$ are constants by definition,   
thus at least one of $l, c$ or $\gm$ must also vary.  
$c$ and $\gm$ cannot vary enough to explain $z$,  
therefore suggesting that $l$ must vary.   
By the variables separable assumption \ref{eq:4.12} 
$l$ is independent of time,  however here it is taken that 
$l$ varies slowly so that \ref{eq:4.12} holds in approximation only.    
In this section $\nu$ is assumed to be the frequency 
of the electromagnetic connecting radiation and the nature of the variation 
in $l$ is taken to be such that $\nu$ maintains the measured value of $z$;  
and this property is a particular example of a property here called the 
{\it preservation of proper frequency}.     
In general at the most fundamental level a co-moving 
quantum system would be expected to have basic states dependent upon a proper 
frequency $\nu$ proportional to $\fr{c^2l(l+2)}{R^2}+\fr{c^4m^2}{\hb^2}$.   
The scale factor $R$ is time dependent,  
in order for the proper frequency to be nearly conserved (or perhaps fixed 
by some other considerations),  it is necessary for $l$ to vary.   
This requirement is here call the {\it preservation of proper frequency}.
This principle might imply that some quantity of matter on microscopic scales
depends on $R$ and $l$.  
Precisely what this quantity may be is unclear.   
The value of the fundamental constants may vary with time,  Dirac (1937) 
\cite{bi:dirac}, and it could be that these depend on $R$ and $l$.   
An alternative way of viewing the preservation of proper frequency 
is by using the Planck equation;   because  $\nu$ remains nearly a constant 
the energy $E=\hb \nu$  will remain nearly constant.
From \ref{eq:2.37} and \ref{eq:2.38} the characteristic time interval 
corresponding to one unit of discrete redshift is
\be
t_{char}=\fr{v_I}{cH_0}=3\pm1\times10^6~~~{\rm years},
\label{eq:5.3}
\ee
thus making direct measurements depending on $l$ unlikely.
Suppose that the value of discrete redshift \ref{eq:2.38} corresponds to $l$ 
varying by a factor of one
\be
\de z=\fr{v_I}{c}=z_{i+1}-z_{i}=\fr{\nu_{i+1}-\nu_i}{\nu_0}.
\label{eq:5.4}
\ee
Using the value of $\nu_0$ from \ref{eq:5.1}
\be
\fr{v_I}{c(1+z)}=\left(\fr{1+4/l+(\fr{9\gm}{4}+\fr{3}{2})/l^2}
  {1+2/l+(\fr{9\gm}{4}-\fr{3}{2})/l^2}\right)^{\fr{1}{2}}-1
\label{eq:5.5}
\ee
expanding for large $l$
\be
\fr{v_I}{c(1+z)}=\fr{1}{l}+O(l^{-2}).
\label{eq:5.6}
\ee
From \ref{eq:2.37} this gives $l\sim 10^7$.   As $l$ is large $l\ap l'$  
and substituting for $l'$ from \ref{eq:5.2}
\be
\fr{v_I\nu_0\sqrt{3\gm-2}}{cH_0}=(2q_0)^{\fr{1}{2-3\gm}}
            \left(2q_0+2-3\gm\right)^{\fr{3\gm}{2(3\gm-2)}}.
\label{eq:5.7}
\ee
This gives a very large $q_0\sim 10^{38}-10^{50}$,  
well outside the observational limits and comparable in size to $10^{42}$ 
the Dirac (1937) \cite{bi:dirac} dimensionless constant.
\subsection{refinements}
The above model has scope for refinements:  for example by using
Maxwell's equations instead of the Klein-Gordon equation,  and more importantly
choosing that the field $\ph$ depends on both $t$ and $\ch$ 
so that the radiation connects observer and emitter;  
however this is not pursued as the model has serious problems which 
it is unlikely that these refinements would overcome.
The most important of these is that it predicts a value for $q_0$
many orders of magnitude larger than allowed for by observation.   
It does not give consistent values for $l$,  if $\nu$ is taken to 
be given by the value given by \ref{eq:2.31},  and \ref{eq:2.34} 
then $l\sim 1$,  \ref{eq:5.2} gives $l>10^{26}$, and \ref{eq:5.6} 
gives $l\sim 10^{7}$.  It predicts that $v_I$ should depend on
frequency and this is not observed,  $v_I$ has the same value using either 
optical or radio data.   The model is not quantum mechanical because
the massless Klein-Gordon equation does not depend explicitly on $\hb$.  
The positive aspects of the model are that it is simple and predicts apparent
discrete motion.
\section{Discrete Redshift via the Massive Klein-Gordon Equation.}
\label{sec:dsvKG}
\subsection{associate Klein-Gordon momenta with the comoving velocity}
The scalar field solutions of section \ref{sec:KG} 
can be interpreted as being the wave function for an element of 
quantized matter in Robertson-Walker spacetime.   
The stress of the scalar field can be calculated by substitution.
Associated with this stress is the momentum density
\be
P_a=T_{ab}U^b,
\label{eq:6.1}
\ee
where $U^b$ is a time-like vector-field tangential to an observers world-line.
In the present case $U^b$ can be taken to be the co-moving vector 
\ref{eq:2.21}.   Corresponding to the momentum density \ref{eq:6.1} 
there is the velocity vector
\be
V_a=\fr{4\pi R^3}{3m}P_a.
\label{eq:6.2}
\ee
In the present case this vector has only the component $V_t$.   
This velocity vector can be taken to be almost co-moving
\be
V_a\approx U_a,
\label{eq:6.3}
\ee
where $U_a$ is the co-moving vector \ref{eq:2.21}.   
Two immediate consequences of this assumption are that:   
the redshift \ref{eq:2.23-25}
constructed using $V_a$
is almost the same as that constructed from $U_a$,  
and that the redshift is discrete.  
The almost co-moving stress assumption can be used to calculate the size of 
the scale factor by methods similar to those of the previous section.   
It predicts a scale factor of microscopic size.
A variant of this approach is to take the vector
\be
W_a=\fr{\ph_a}{\sqrt{-\ph_c\ph^{\dag c}}},
\label{eq:6.4}
\ee
to be almost co-moving.   
This variant is similar to the co-moving stress approach,  but it also 
predicts angular terms because of non-vanishing of $W_i$.
The reason that the moving stress approach gives the wrong result is that 
it is unrealistic to consider the stress given by a single scalar field.
It might be hoped that a statistical ensemble of such wave functions would 
produce the correct co-moving stress.   It is not immediately apparent how
how to construct such ensembles.   For a review of statistical mechanics in
curved spacetime see Ehlers (1971) \cite{bi:ehlers}.   
It can be assumed that such statistical 
ensembles give back the solutions of section  \ref{sec:intro} 
and that the single 
Klein-Gordon equation produces a quantum perturbation of this.   
This approach is pursued in section  \ref{sec:dsvQP}.
\subsection{other massive Klein-Gordon approaches}
    There are several other approaches based on solutions to the massive
Klein-Gordon equation.   In non-relativistic quantum mechanics there is the 
equation
\be
<p_a>=-i\hb\int dx^a\ph^\dag\p_a\ph,
\label{eq:6.5}
\ee
see for example Schiff (1949) \cite{bi:schiff} equation 7.8.   
This equation would give non-vanishing $p_i$ components,  
so that the requirement that $\fr{p_a}{m}$ 
is approximately co-moving again would not hold;  
also it is not clear what the interpretation of \ref{eq:6.5} 
is in relativistic quantum mechanics.
Another approach is to use the group velocity $c_{group}$ of $\ph$,  
as defined by Schr\"odinger (1939) \cite{bi:schro39},  
to define an effective energy and hence $p_t$ component;
however the existence of a non-negligible group velocity $c_{group}$
would again imply non-negligible $p_i$  components.   
There is an entirely different approach which
consists of investigating how the hydrogen atom behaves in a Robertson-Walker 
universe,  see for example Trees (1956) \cite{bi:trees};  
the problem with this approach is that it requires the value of $l$ 
to be small and is  thus unsuitable for the problem in hand.
\section{Discrete Redshift via Quantum Perturbation of Classical Solutions.}
\label{sec:dsvQP}
\subsection{general weak metric perturbation}
The metric is taken to be of the form
\be
g_{ab}=\bar{g}_{ab}+h_{ab},~~~g^{ab}=\bar{g}^{ab}-h^{ab},
\label{eq:7.1}
\ee
where $\bar{g}_{ab}$ is a given background field,  
in the present case this is the Robertson-Walker metric \ref{eq:2.1},  
and $h_{ab}$ is a small perturbative term.   The connection is
\be
\Ga^a_{bc}\equiv\{^a_{bc}\}+\fr{1}{2}K^{a}_{.bc},
\label{eq:7.2}
\ee
where the contorsion is
\be
K_{abc}=h_{ba;c}+h_{ca;b}-h_{bc;a},
\label{eq:7.3}
\ee
and $\{^a_{bc}\}$ is the Christoffel symbol 
of the background field $\bar{g}_{ab}$,
$";"$  is the covariant derivative with respect 
to the background field $\bar{g}_{ab}$.
For any connection which is a sum of the Christoffel connection and a 
contorsion tensor,  the Riemann tensor is
\be
R^a_{.bcd}=\bar{R}^a_{.bcd}+K^a_{.~[d|b|;c]}
           +K^a_{.eb}S^e_{.cd}+\fr{1}{2}K^a_{.~[c|e|}K^e_{.~d]b}.
\label{eq:7.4}
\ee
In the present case the torsion $S^a_{.bc}$ and the cross terms in $h_{ab}$ 
are taken to vanish,  then after using \ref{eq:1.3} 
for the commutation of covariant derivatives,  the Riemann tensor becomes
\be
R^a_{.bcd}=\bar{R}^a_{.bcd}-\fr{1}{2}h^a_{.e}\bar{R}^e_{.bcd}
                           -\fr{1}{2}h_{be}\bar{R}^{ea}_{..cd}
 +\fr{1}{2}(h^{~a}_{d.~;bc}-h^{~~~a}_{db;.~c}
           -h^{~a}_{c.~;bd}+h^{~~~a}_{cb;.~d}),
\label{eq:7.5}
\ee
Contracting and again using \ref{eq:1.3}
\be
R_{bd}=\bar{R}_{bd}-h^{fe}\bar{R} _{ebfd}
      +\fr{1}{2}h_{be}\bar{R}^e_{.d}+\fr{1}{2}h_{de}\bar{R}^e_{.b}
 +\fr{1}{2}(h^{~c}_{d.,cb}+h^{~c}_{b.;cd}-\Box h_{db}-h_{;bd}),
\label{eq:7.6}
\ee
where $h=h^a_{.a}$.   Equations \ref{eq:7.4},  \ref{eq:7.5},  and \ref{eq:7.6}
differ from (I4) and (I5) of Lifshitz and Khalatnikov (1963) \cite{bi:LK}
as they leave out all the terms involving products of 
$h_{ab}$ and $\bar{R}_{cdef}$.
\subsection{perturbing the stress of a perfect fluid}
Perturbations of the stress usually (see for example Lifshitz and
Khalatnikov (1963) \cite{bi:LK} and Sacks and Wolfe (1967) \cite{bi:SW}) 
are of a perfect fluid which obeys
\be
R_{ab}=(\mu+p)U_aU_b+\fr{1}{2}(\mu-p)g_{ab},
\label{eq:7.7}
\ee
the first conservation equation
\be
\mu_aU^a+(\mu+p)U^a_{.;a}=0,
\label{eq:7.8}
\ee
and the second conservation equation
\be
(\mu+p)\dot{U}_a +(g_a^b+U_aU^b)p_{,b}=0,
\label{eq:7.9}
\ee
where
\be
\dot{U}_a=U_{a;b}U^a.
\label{eq:7.10}
\ee
The perfect fluid is linearly perturbed thus
\be
\mu=\bar{\mu}+\de\mu,~~~
p=\bar{p}+\de p,~~~  
U_a=\bar{U}_a +\de U_a,~~~  
g_{ab}=\bar{g}_{ab}+\de g_{ab}.
\label{eq:7.11}
\ee
Identifying
\be
\de\mu=-\ph_c\ph^c+V(\ph),~~~  
\de p=-\ph_c\ph^c-V(\ph),~~~
\de U_a=\fr{\ph_a}{\sqrt{-\ph_c\ph^c}},~~~
\de g_{ab}=h_{ab},
\label{eq:7.12}
\ee
the Klein-Gordon equation and the scalar field stress are recovered at
the second and third orders.   The solution to the Klein-Gordon equation
given in Section \ref{sec:KG} are not compatible with this linearization 
because the first order perturbation produces cross equations in the perfect 
fluid and scalar field which are not obeyed.   This can be readily verified by
investigating the time component of the first order perturbation of the
second conservation equation \ref{eq:7.9} for the Einstein static universe 
this gives
\be 
\left(\fr{\ph_t}{\sqrt{-\ph_c\ph^c}}\right)_{,t}=0,
\label{eq:7.13}
\ee
which is incompatible with the solution \ref{eq:4.15}.   
Usually perturbation theory fixes the values of $\de \mu$ and $\de p$,  
thus not giving the freedom necessary to replace them with scalar fields.
\subsection{perturbing the stress of a scalar field}
Here perturbation are taken to given by the scalar field stress 
\ref{eq:4.28}, thus
\be
R_{ab}=\bar{R}_{ab}+\ph_a\ph^\dag_b+\ph^\dag_a\ph_b
      +g_{ab}\fr{m^2c^2}{\hb^2}\ph\ph^\dag,
\label{eq:7.14}
\ee 
The components of the Ricci tensor follow immediately after noting
\ber
\ph\ph^\dag&=&A^2YY^\dag\left(C_+^2+C_-^2
                              +2C_+C_-\cos(2\bar{\nu})\right)\nonumber\\
\ph_t\ph_t^\dag&=&A_t^2YY^\dag\left(C_+^2+C_-^2
                              +2C_+C_-\cos(2\bar{\nu})\right)\nonumber\\
               &&-4AA_tYY^\dag \sin(2\bar{\nu})\nonumber\\
               &&+A^2YY^\dag\left(C_+^2+C_-^2
                              -2C_+C_-\cos(2\bar{\nu})\right)\nonumber\\
\ph_i\ph^\dag_t+\ph_i^\dag\ph_t&=&AA_t(YY_i^\dag+Y_iY^\dag)\left(C_+^2+C_-^2
                              +2C_+C_-\cos(2\bar{\nu})\right)\nonumber\\
    &&+iA^2(YY_i^\dag-Y_iY^\dag)(C_+^2-C_-^2)\bar{\nu}_t\nonumber\\
        &&-2(YY_i^\dag+Y_iY^\dag)C_+C_-\bar{\nu}_t\sin(2\bar{\nu}),\nonumber\\
\ph_i\ph_j^\dag+\ph_i^\dag\ph_i&=&A^2(Y_iY_j^\dag+Y_i^\dag Y_j)
      \left(C_+^2+C_-^2+2C_+C_-\cos(2\bar{\nu})\right).
\label{eq:7.15}
\ear
\subsection{the $R_{tt}$ component}
In general the equations resulting from \ref{eq:7.14} are intractable  and
therefore attention is restricted to the Einstein static universe.   The
perturbations are taken to be in the harmonic gauge
\be
h^{~b}_{a.;b}=\fr{1}{2}h_{,a},
\label{eq:7.16}
\ee
the $R_{tt}$ component of \ref{eq:7.14} is
\ber
R_{tt}-\bar{R}_{tt}&=&-\fr{1}{2}\Box h_{tt}\nonumber\\
                   &=&YY^\dag\left\{(C_+^2+C_-^2)
   (2\nu^2-\fr{c^2m^2}{\hb^2}(c^2-h_{tt})\right.\nn
                   &&\left.+2C_+C_-\cos(2\nu t)
   (-2\nu^2-\fr{c^2m^2}{\hb^2}(c^2-h_{tt}))\right\}.
\label{eq:7.17}
\ear
due to the presence of $YY^\dag$,  which obeys \ref{eq:4.9},  
this equation appears to be intractable.   
Now $YY^\dag=\sin^{|2n|}(\al)\cos^{|2n|}(\al)$,   
expanding the trigonometrical 
functions and taking $n=0$ (implying $p=l$) gives $YY^\dag$ 
is approximately one for small $\al$.   
Taking the cross term $m^2h_{tt}$ to be negligible and assuming 
that $h_{tt}$ is only a function of $t$,  \ref{eq:7.17} reduces to
\ber
h_{tt,tt}&=&2C^{-2}(C_+^2+C_-^2)\left(2\nu^2-\fr{c^4m^2}{\hb^2}\right)\nn
        &&-4C^{-2}C_+C_-\left(2\nu^2+\fr{c^4m^2}{\hb^2}\right)\cos(2\nu t),
\label{eq:7.18}
\ear
which has solution
\ber
h_{tt}&=&C_{-2}(C_+^2+C_-^2)\left(2\nu^2-\fr{c^4m^2}{\hb^2}\right)t^2\nn
      &&+C_+C_-\fr{2\nu^2+c^4m^2\hb^{-2}}{c^2\nu^2}\cos(2\nu t)+B_1+B_2,
\label{eq:7.19}
\ear
where $B_1$ and $B_2$ are constants.
\subsection{the $R_{ti}$ component}
   The $R_{ti}$ component of \ref{eq:7.14} is
\ber
R_{ti}-\bar{R}_{ti}&=&h_{ti}-\fr{1}{2}\Box h_{ti}\nonumber\\
                   &=&-2(YY_i^\dag+Y_iY^\dag)C_+C_-\nu \sin(2\nu t)\nonumber\\ 
                   &&+i(YY_i^\dag-Y_iY^\dag)(C_+^2-C_-^2)\nu\nonumber\\
          &&+\fr{c^2m^2}{\hb^2}h_{ti}YY^\dag(C_+^2+C_-^2+2C_+C_-\cos(2\nu t)).
\label{eq:7.20}
\ear
The $i=\bt$ and $i=\ph$ components of the first term vanish,  
however the $i=\al$ term does not as 
$YY_\al^\dag+Y_\al Y^\dag=2(n \cot(\al)- p \tan(\al))YY^\dag$;  
again taking $n=0$ and expanding the trigonometrical functions 
gives that this term vanishes for small $\al$.
By equations \ref{eq:4.11} the angular part of the second term is non-vanishing
and remains so after expanding for small $\al$,  by assuming that $C_+^2$ 
is of the same magnitude as $C_-^2$ this term can be taken to vanish.   
Similarly to the $R_{tt}$ component the $m^2h_{ti}$ term can be taken 
to be negligible.   Hence all of the left hand side of \ref{eq:7.20} 
can be taken to vanish,  thus $h_{ti}=0$ is an approximate solution 
to \ref{eq:7.20}.
\subsection{the $R_{ij}$ component}
The $R_{ij}$ component of \ref{eq:7.14} is
\ber
R_{ij}-\bar{R}_{ij}&=&-g_{ij}h^k_{.k}R_0^{-2}+3h_{ij}
                      -\fr{1}{2}h_{ij}\nonumber\\
 &=&\left(Y_iY_j^\dag+Y_i^\dag Y_i^\dag+(\bar{g}_{ij}+h_{ij})
                      \fr{m^2c^2}{\hb^2}YY^\dag\right)\nonumber\\
                &&\times\left(C_+^2+C_-^2+2C_+C_-\cos(2\nu t)\right).
\label{eq:7.21}
\ear
Now $Y_iY_j^\dag+Y_i^\dag Y_j$ is non-vanishing for $i=j=\al,\bt,\ph$ 
and $i=\bt$,  $j=\ph$ or $i=\ph$,  $j=\bt$.
Taking $n=0$ and expanding there remains just the $i=j=\ph$ component and it
is of size $l^2$;  the spatial axes can be rotated so that
\be
Y_iY_j^\dag+Y_iY_j^\dag=l^2\bar{g}_{ij}^{(3)}.
\label{eq:7.22}
\ee
Similarly to the $R_{tt}$ component $YY^\dag$ is taken to be approximately 
one and $m^2h_{ij}$ is taken to be negligible.  Subject to the ansatz
\be
h_{ij}\equiv\si(t)\bar{g}_{ij}^{(3)},
\label{eq:7.23}
\ee
\ref{eq:7.21} reduces to
\be
\si_{,tt}=2c^2\left(l^2+\fr{m^2c^2R_0^2}{\hb^2}\right)
           \left(C_+^2+C_-^2+2C_+C_-\cos(2\nu t)\right),
\label{eq:7.24}
\ee
which has solution
\ber  
\si&=&+c^2(C_+^2+C_-^2)\left(l^2+\fr{c^2m^2R_0^2}{\hb^2}\right)l^2\nn
    &&-c^2C_+C_-\fr{l^2+c^2m^2R_0^2\hb^{-2}}{\nu^2}\cos(2\nu t) 
    +B_3 t+B_4,
\label{eq:7.25}
\ear
where $B_3$ and $B_4$ are constants.
\subsection{the harmonic gauge condition}
The solutions for $h_{ab}$ given by \ref{eq:7.19},  \ref{eq:7.23},   
and \ref{eq:7.25} do not obey the harmonic gauge condition \ref{eq:7.16}.   
This is a result of the approximations made for $Y$.   From \ref{eq:7.22} 
$h_{,t}$ will depend on $l$ but from \ref{eq:7.19} $h^{~t}_{t.;t}$ will not,
thus violating the time component of the harmonic gauge condition.   
No approximation for $Y$ which allows the harmonic gauge condition 
to be preserved are known.   \ref{eq:7.19},  \ref{eq:7.23},  
and \ref{eq:7.25} give the weak field metric perturbations
\be
N^2=\bar{N}^2-c^{-2}h_{tt},~~~  
R^2=\bar{R}^2+\si.
\label{eq:7.26}
\ee
Note that the equations \ref{eq:7.17}, \ref{eq:7.20}  and \ref{eq:7.21} 
are not equivalent to the differential equations that arise if 
the substitutions
\be
N=\bar{N}+\ep_1,~~~          
R=\bar{R}+\ep_2,
\label{eq:7.27}
\ee
are used in \ref{eq:2.10} and \ref{eq:2.11},  because for example,  
there are no second derivatives of $N$ in \ref{eq:2.10} and \ref{eq:2.11}.
\subsection{the change in redshift}
In general weak metric perturbations induce a complicated change in the 
redshift.   This has been calculated for the conformally flat ($k=0$) 
case by Sacks and Wolfe (1967) \cite{bi:SW}.   
The present case is much simplified because the metric perturbations 
are of the form \ref{eq:7.26},  the new values of $N$ and $R$ can be
used for the vectors \ref{eq:2.31} and \ref{eq:2.34} to give the redshift 
of the form \ref{eq:2.36},  thus
\be
1+z=\fr{R_0}{R}=\fr{R_0}{\bar{R}}\left(1+\fr{\si}{\bar{R}^2}\right)
   \ap\fr{R_0}{\bar{R}}\left(1-\fr{\si}{2\bar{R}^2}\right).
\label{eq:7.28}
\ee
Appealing to the principle of the preservation of proper frequency,  introduced
in section \ref{sec:dsvCR},  a change in the value of $l$ by a factor of one 
is taken to correspond to one unit of discrete redshift
\be
\fr{v_I}{c}=|\de z|,~~~\de z\equiv z_{l+1}-z_l
           \ap\fr{1+z}{2\bar{R}^2}(\si_l-\si_{l+1}).
\label{eq:7.29}
\ee
This equation allows rough estimates to be made of the size of $R_0$;  
as such an estimate can only be made of the order of magnitude of $R$ 
it can be assumed that $\fr{1+z}{2\bar{R}^2}\ap R_0^2$,  
this still leaves $C_+,  C_-,  l,  t$ and $R$ of unknown size.   
The frequency $\nu$ is large compared to the time scales involved in
\ref{eq:7.25}, as for example the Compton frequency of the electron
$\nu_e=\fr{m_ec^2}{\hb}\ap10^{21}~{\rm sec}^{-1}$;
thus the $t^2$ term in \ref{eq:7.25} is larger than the $\cos$ term 
and \ref{eq:7.29} becomes
\be
R_0^2=(C_+^2+C_-^2)(2l+1)t^2c^3v_I^{-1}.
\label{eq:7.30}
\ee
Using the $T^t_{.t}$ component of the scalar field's stress
\be
C_+^2+C_-^2=\fr{8\pi G}{c^2}
             \left(\fr{2l}{R_0^2}+\fr{c^2m^2}{\hb^2}\right)^{-1},
\label{eq:7.31}
\ee
where $\mu_S$ is the density of the scalar field,  
and the $\cos$ term is taken to be negligible and $YY^\dag\sim 1$,  
thus \ref{eq:7.30} becomes
\be
R_0^2=8\pi G\mu_S\left(\fr{2l}{R_0^2}+\fr{c^2m^2}{\hb^2}\right)
       (2l+1)t^2\fr{c}{v_I}.
\label{eq:7.32}
\ee
It has been assumed that $C_+^2+C_-^2$ is a constant and that $\mu_S$ is $l$ 
dependent,  this implies that the substitution for $C_+^2+C_-^2$ 
takes place after equation \ref{eq:7.29} has been applied.
\subsection{some incompatible conditions}
Rather than deriving a value of $R_0$ from \ref{eq:7.32},  it is shown what 
values of $R_0,  l,  t$ and $s$ are compatible with this equation.   
First it is proved that the following conditions are incompatible:
\newline
i)\ref{eq:7.32} holds,\newline
ii) the $"t"$ in \ref{eq:7.32} is less than $H_0^{-1}$,\newline
iii)$R>10^{22}$ meters (this is a typical distance between galaxies),\newline
iv)$\mu_s<10^5{\rm Kg}~ {\rm m}^{-3}$ (this is an extremely high density 
compared with\newline
$~~~~~~~~~~~~~~~~~~~~~~~~~~~~$ a typical stellar interior density)
\newline   
$~~~\mu_c=10^{-26}~{\rm Kg}~ {\rm m}^{-3}$  the critical density for a 
                               Robertson-Walker universe,\newline
$~~~~\mu=10^{-5}~~{\rm Kg}~ {\rm m}^{-3}$ a typical photosphere density 
                                (see p.163 Allen (1973)),\newline
$~~~\mu=10^{+3}~~{\rm Kg}~ {\rm m}^{-3}$ an average stellar density;   
it would be expected that $\mu_s<\mu_c$ for the weak metric approximations 
used in deriving \ref{eq:7.32} to work),\newline
v)$\nu_{geomerty}<<\nu_e$ (this is necessary if the scalar field $\ph$ 
is chosen to represent a known field).\newline
Proof:  v) implies that the $\fr{l}{R}$ term in \ref{eq:7.32} 
can be neglected thus
\be
R_0^2=\fr{8\pi G}{c^2}\fr{\hb^2}{m^2c^2}(2l+1)\fr{t^2c^2}{v_I},
\label{eq:7.33}
\ee
using ii)
\be
R_0^2<\fr{8\pi G\mu_S\hb^2(2l+1)}{m^2cv_IH_0^2},
\label{eq:7.34}
\ee
again using v)
\be
R_0<\fr{16\pi G\mu_s \hb}{mv_IH_0^2}, 
\label{eq:7.35}
\ee
in $SI$ units this is $R_0<10^{18}\mu_S$,  
from which conditions ii) and iii) can be seen to be incompatible.
\subsection{some compatible conditions}
The most realistic compatible conditions are:\newline
i) \ref{eq:7.32} holds,\newline
ii) the $"t"$ in \ref{eq:7.32} equals $H_0^{-1}$,\newline
iii) $R_0=10^{28}$ meters $~~$(this implies that $2q_0-1=10^{-4}$),\newline
iv) $\mu_S=10^{-13}{\rm Kg}~ {\rm m}^{-3}$ 
(this is well above the critical density but below\newline
$~~~~~~~~~~~~~~~~~~~~~~~~~~~~~~~~~$ a typical photosphere density),\newline
v) $\nu=t_p^{-1}$,  where 
$t_p^{-1}=\hb^{\fr{1}{2}}G^{-\fr{1}{2}}c^{-\fr{5}{2}}\ap10^{-44}{\rm sec}.$ 
is the Planck time,\newline
$~~~~~~~~~~~~~~~~~~~~~~~$  (this forces the scalar field 
$\ph$ to be a hypothetical field rather than a known field,\newline
$~~~~~~~~~~~~~~~~~~~~~~~$   together with the above value of $R_0$  
it implies that $l=10^{63}$ which is the Dirac\newline
$~~~~~~~~~~~~~~~~~~~~~~~$   dimensionless constant 
                            to the power of $\fr{3}{2}$).\newline
Proof:  Re-arranging \ref{eq:7.32}
\be
\fr{c^2m^2R_0^2}{\hb^2}=2l(8\pi G\mu_St^2cv_I^{-1}-1)+8\pi G\mu_St^2cv_I^{-1},
\label{eq:7.36}
\ee
using ii) for the value of $t$ and using $SI$ units
\be
10^{25}R_0^2\ap2l(10^{31}-1)+10^{31}.
\label{eq:7.37}
\ee
Now v) implies that $\nu_{geometry}>>\nu_{compton}$ giving
\be
l\ap10^{35}R_0\ap10^{-6}\mu_S^{-1}R_0,
\label{eq:7.38}
\ee
and the values of $R_0  ,\mu_S$  and $l$ given above can be shown 
to obey \ref{eq:7.38} by substitution.
\subsection{summary}
To summarize some of the deficiencies of the above model.   There are at 
least three  technical deficiencies:  there is no proof that the dynamical 
equations \ref{eq:7.14} are consistent;  the Einstein static universe has been
assumed in order to solve the perturbation equations,  but the background 
metric is time dependent;  and various approximations have been made for the
angular terms,  in particular the approximations \ref{eq:7.22} 
result in the loss of the initially assumed harmonic gauge condition.   
There are at least three physical deficiencies:  
the equation for discrete redshift in the form \ref{eq:7.29}
depends on $z$ contrary to observation;   the most realistic compatible 
conditions for equation \ref{eq:7.32} require a hypothetical field with the 
unusual property of a frequency of the order of the inverse Planck time;  
and the result requires a scalar field density above the critical density 
of a Robertson-Walker-Friedman universe.
\section{Conclusion.}
\label{sec:conc}
Properties of Robertson-Walker spacetimes can be discrete if they depend on
the spherical harmonic integer $l$,  in particular redshift is discrete 
if there is a mechanism to connect it to this integer.   
Density perturbations have been known for a long time 
to depend on this integer and thus integer dependent 
redshift could have been predicted before it was observed.   
The problems with introducing discrete redshift via density 
perturbations include:  
density perturbations are irregular whereas the value of discrete redshift 
is constant irrespective of other conditions,  
and more importantly density perturbations 
provide no mechanism which will alter $l$.  
Solutions to the Klein-Gordon equation and other field equations 
also depend on the spherical harmonic integer $l$.
The requirement that the energy of these fields is almost conserved implies 
that,  as discussed in section \ref{sec:dsvCR},  
the proper frequency is preserved;  this in 
turn implies that the value of $l$ changes in a regular manner 
proportional to the increase in the scale factor $R$.   
In principle all quantum fields have the Universe as an ultimate boundary 
condition and are thus presumably $l$ dependent.
Here it was found that theories using only solutions to the Klein-Gordon 
equation predict a microscopic value for the scale factor $R$.   
It was suggested that the large scale behaviour of Robertson-Walker 
spacetime is governed,  as it is classically,  by the Friedmann equation 
and that the Klein-Gordon solutions in this background induce weak metric 
perturbations of the spacetime.   It might be that solutions for fields 
involve other integers,  apart from the spherical harmonic integer,  
and this could also lead to discrete redshift via 
induced metric perturbations.   From \ref{eq:7.12} it is not clear that 
this coupling is well-defined as the interactions between the Klein-Gordon 
field and the background fluid may be non-negligible.   
To produce a realistic prediction of the size of the 
scale factor using induced metric perturbations,  
it was necessary to make some very coarse technical and physical assumptions,  
including the requirement that the scalar field has a frequency approximately
equal to the inverse Planck time,  this precludes the scalar field representing
a known particle.   Any theory based upon metric perturbations will predict
what in section \ref{sec:remarks} is called real discrete motion;   
this implies that there should be boundary effects where the value of 
the discrete redshift jumps.
It is hoped that using the theory of quantum fields on curved spacetimes will
remove,  or reduce the bounds on,  the free parameters 
such as $C_+$ and $C_-$
in the induced weak metric perturbations,  
and will give a more rigorously defined theory.
\footnote{The referee suggests:'- take away ``$2q_0-1=10^{-4}$'' 
from the Abstract and add in the Conclusions that the value 
$q_0=1/2$ is obtained in the absence of the cosmological constant.   
Taking it into account, the value of $q_0$ would be in agreement 
with the recent observations on the Ia type supernovae.'}
\section{Acknowledgement}
I would like to thank Tony Fairall and Tim Gebbie
for reading and commenting on this paper..
This work was supported in part by the South African Foundation
for Research and Development (FRD).

\end{document}